\documentclass[12pt]{article}

\usepackage{color}
\usepackage[dvips]{graphicx,psfrag}
\usepackage{amsmath,amssymb}
\usepackage{bm}

\baselineskip=\normalbaselineskip
\renewcommand{\baselinestretch}{1.34}
\renewcommand{\thefootnote}{\fnsymbol{footnote}}

\newcommand{\ket}[1]{\bigl|#1\bigr>}
\newcommand{\bra}[1]{\bigl<#1\bigr|}
\newcommand{\bracket}[2]{\left.\left\langle #1\right|#2\right\rangle}
\newcommand{\maru}[1]
{{\ooalign{\hfil#1\/\hfil\crcr\raise.167ex\hbox{\mathhexbox20D}}}}

\newcommand{\rmq}{{\rm q}}

\newcommand{\bP}{\boldsymbol{P}}

\newcommand{\bunit}{\boldsymbol{1}}

\newcommand{\bL}{\boldsymbol{L}}

\newcommand{\bsigma}{\boldsymbol{\sigma}}

\newcommand{\eq}[1]{(\ref{#1})}

\newcommand{\nn}{\nonumber}

\DeclareMathOperator{\tr}{tr}

\DeclareMathOperator{\sgn}{sgn}

\allowdisplaybreaks[1]

\makeatletter
\@addtoreset{equation}{section}
\@addtoreset{theorem}{section}
\@addtoreset{definition}{section}
\@addtoreset{lemma}{section}
\@addtoreset{proposition}{section}

\begin{document}
\begin{flushright}
\parbox{40mm}{%
KUNS-2072 \\
{\tt arXiv:0706.4471} \\
June 2007}
\end{flushright}

\vfill

\begin{center}
{\Large{\bf 
Notes on D-branes and dualities \\
in $(p,q)$ minimal superstring theory
}}
\end{center}

\vfill

\begin{center}
{\large{Hirotaka Irie}}\footnote%
{E-mail: {\tt irie@gauge.scphys.kyoto-u.ac.jp}}   \\[2em]
Department of Physics, Kyoto University, 
Kyoto 606-8502, Japan \\

\end{center}
\vfill
\renewcommand{\thefootnote}{\arabic{footnote}}
\setcounter{footnote}{0}
\addtocounter{page}{1}

\begin{center}
{\bf abstract}
\end{center}

\begin{quote}
We study boundary states 
in $(p,q)$ minimal superstring theory, combining 
the explicit form of matter wave functions. 
Within the modular bootstrap framework, 
Cardy states of $(p,q)$ minimal superconformal field theory 
are completely determined 
in both cases of the different supercharge combinations, 
and the remaining consistency checks 
in the super-Liouville case are also performed. 
Using these boundary states, we determine the explicit 
form of FZZT- and ZZ-brane boundary states both 
in type 0A and 0B GSO projections. Annulus amplitudes 
of FZZT branes are evaluated and 
principal FZZT branes are identified. 
In particular, we found that 
these principal FZZT branes do not satisfy 
Cardy's consistency conditions 
for each other and play a role of order/disorder 
parameters of the Kramers-Wannier duality 
in spacetime of this superstring theory. 
\end{quote}
\vfill
\renewcommand{\baselinestretch}{1.4}
\newpage

\section{Introduction and summary}

Noncritical (super)string theories 
have been investigated as a useful laboratory
of critical string theory 
and have been discussed in various contexts: 
the worldsheet description 
\cite{Polyakov,KPZ,DHK,Seiberg,DOzz,doZZ,tes1, sDOZZ,
fzz-t,zz,tes2,fuku-hoso,ars,Mar,SeSh,KOPSS,Oku,bdyGR,recentsL}, 
matrix models 
\cite{dsl,Kostov,mss,ms,dkk,2cut1MM,ComMM,tt,newhat,UniCom,MMSS,grt,SeSh2} 
and the string field theoretical descriptions \cite{fy1,fy2,fy3,fis,fim,fi1,fi2}. 
They have fewer degrees of freedom and 
we can make a detail study
of many important properties shared 
with their critical counterparts. 
Moreover, since these string theories are described with many 
different formulations, 
we can investigate various aspects of stringy phenomena. 

Here we further study $(p,q)$ minimal superstring theory 
from the conformal field theory (or Liouville theory) approach. 
Minimal superstring theory is 
one of the most tractable superstring theories 
among these noncritical superstring theories. 
Its worldsheet description is defined with 
$\mathcal N=1$ super Liouville field theory \cite{Polyakov,KPZ,DHK} 
coupled to $\mathcal N=1$ $(p,q)$ minimal superconformal field theory 
\cite{bpz,fms,smini,corsmini} 
with type 0 GSO projection (see \cite{UniCom,SeSh} 
for its basic properties). 
From the recent developments in Liouville field theory 
without boundaries \cite{DOzz,doZZ,tes1,sDOZZ} 
and with boundaries \cite{fzz-t,zz,tes2},
the boundary states of super Liouville field theory 
were obtained in \cite{fuku-hoso,ars}, 
and in the framework of minimal superstring theory 
they have been extensively studied in \cite{SeSh,SeSh2,Oku}, where 
the disk amplitudes of corresponding D-branes 
were explicitly evaluated \cite{SeSh} 
and the pure-supergravity case, $(p,q)=(2,4)$, 
was studied including annulus amplitudes \cite{SeSh2,Oku}. 
For further investigation of these D-branes,
however, 
we need to know all the  Cardy states 
of $(p,q)$ minimal superconformal field theory 
with different supercharge combinations of left and right: 
\begin{align}
\bigl(G_r-i\eta\, \bar G_{-r}\bigr)\ket{B;\eta}=0 
\qquad (\eta=\pm 1), \label{scibc}
\end{align}
and we need to combine 
Cardy states of each SCFT in practice. 

As is known from the original works on 
the conformally invariant boundary states 
\cite{cardy1,Ishibashi,cardy3}, Cardy states have 
a one-to-one correspondence with the highest weight 
representations of the open-channel Virasoro algebra.
As is briefly reviewed in section 2.1, 
Cardy states have the following form 
in this superconformal case:%
\footnote{
Since open strings between opposite $\eta$ boundaries are 
in Ramond sector, we make use of the convention that 
Cardy states of $\eta=-1 \ (+1)$ are labeled with 
the highest weight in the NS (R) sector of 
the open-string super-Virasoro algebra. 
}
\begin{align}
\ket{h_{NS\pm}} 
 &=\sum_{i\in NS}
     B_{h_{NS}}{}^{i_{NS}}
     \ket{i_{NSNS};\eta=-1}\bigr>
  \pm\sum_{i\in R} 
     B_{h_{\widetilde{NS}}}{}^{i_R}
     \ket{i_{RR};\eta=-1}\bigr>,  \nn\\
\ket{h_{R\pm}} 
 &=\sum_{i\in NS}
     B_{h_{R}}{}^{i_{\widetilde{NS}}}
     \ket{i_{NSNS};\eta=+1}\bigr>
  \pm\sum_{i\in {\widetilde R}} 
     B_{h_{\widetilde{R}}}{}^{i_{\widetilde{R}}}
     \ket{i_{RR};\eta=+1}\bigr>.
\end{align}
We often use a notation 
like $j_{RR}$ in this paper 
to indicate the sector to which 
an index $j$ belongs, and 
$\ket{j;\eta}\bigr>$ is a corresponding Ishibashi state \cite{Ishibashi}. 
Three kinds of these wave functions $B_h{}^i$ ($h$ and $i$ are among NS, 
${\rm \widetilde{NS}}$ and R sectors) 
are known from the work of \cite{SCardy},%
\footnote{
The Cardy states of Ramond ground states 
(denoted as $\ket{\theta_{R\pm}}$ in this paper) 
are not considered in \cite{SCardy}. 
This means that only odd models are considered. 
In this paper, we also derive the formula of this case.}
and they are written with the modular 
matrices $S_h{}^i$ of this theory \cite{modSsmini} as
\begin{align}
B_{h_{NS}}{}^{i_{NS}} 
=\frac{1}{\sqrt{2}}\frac{S_{h_{NS}}{}^{i_{NS}}}
{\sqrt{S_{0_{NS}}{}^{i_{NS}}}}, \quad
B_{h_{\widetilde{NS}}}{}^{i_{R}} 
=\frac{1}{2^{3/4}}
\frac{S_{h_{\widetilde{NS}}}{}^{i_{R}}}
{\sqrt{S_{0_{\widetilde{NS}}}{}^{i_{R}}}}, \quad
B_{i_{R}} {}^{h_{\widetilde{NS}}}
=
\frac{S_{i_{R}} {}^{h_{\widetilde{NS}}}}
{\sqrt{S_{0_{NS}}{}^{i_{NS}}}}. \label{wf}
\end{align}
On the other hand, the form of the remaining 
wave functions 
$B_{h_{\widetilde R}}{}^{i_{\widetilde R}}$
is still not known. 
One of the reasons why these wave functions 
$B_{h_{\widetilde R}}{}^{i_{\widetilde R}}$ have not been obtained
is that they do not satisfy such a simple relation with 
modular matrices $S_h{}^i$ like \eq{wf}. Since 
any corresponding characters 
(those in ${\rm \widetilde R}$ sector) 
always vanish 
due to the fermion zero-modes in the cylindrical geometry
(except for the Ramond ground states), 
there is not such 
a modular matrix $S_h{}^i$ 
among these ${\rm \widetilde R}$ characters. 

If one considers 
$(p,q)$ minimal superconformal field theory 
as only a single SCFT, 
this does not cause any problems in studying the boundary 
states as in the literature \cite{bdysmini}. 
It is because we impose the spin-model GSO projection \cite{fms} 
($\Gamma=-1$ on RR states) 
and the above remaining states are always projected out. 
On the other hand, the case of the superstring theory is not so. 
Since superstring theory is 
a combined superconformal field theory (Liouville, matters and ghosts), 
the type 0 GSO projection cannot 
eliminate these contributions \cite{UniCom,SeSh}. 
So this is the first task of this paper: 
We will completely determine all the wave functions of Cardy states, 
including the remaining wave functions, 
in the case of $(p,q)$ minimal superconformal field theory (section 2.2). 

Actually, the way to obtain these remaining wave functions is very simple. 
All we have to do is to consider the OPE algebra with the simplest 
degenerate primaries $(1,2)_+$ \cite{bpz} in the sense of open (or chiral) 
superconformal field theory 
and to consider the Cardy equations obtained 
from the inner products with the Ramond Cardy state 
$\ket{(1,2)_{\rm R+}}$. 
The plus symbol in the subscript of $(1,2)_+$ 
means that this is a primary operator of positive chirality. 
Essentially we use the following fusion rule that could be read off 
from the super Coulomb gas formalism in Ramond sector \cite{corsmini}:
\begin{align}
\mathcal N_{(1,2)_+,(k,l)_+}{}^{(r,s)}
=\delta_{(k,l+1)_+}{}^{(r,s)} 
+\delta_{(k,l-1)_-}{}^{(r,s)},
\end{align}
where operators of both positive and negative chiralities 
come into the relations. 
This kind of fusion rule is also found 
in the super-Liouville cases \cite{fuku-hoso} and we actually show that 
all of the results obtained within 
the conformal bootstrap methods \cite{fuku-hoso} 
are consistent with our procedure. 
This means that all the information about boundary states 
can be extracted from the modular bootstrap methods
also in the case of $\mathcal N=1$ superconformal field theory.%
\footnote{Of cause, we need to perform the conformal bootstrap method to 
know some relations with variables in the boundary action (e.g., boundary 
cosmological constants $\zeta$ in Liouville theory). }

In section 2.3, we present the complete form of the boundary states 
in both cases of 0B and 0A theory, and 
investigate their basic properties. 
As in the case of 
super Liouville field theory \cite{fuku-hoso}, 
the wave functions of $\eta=-1/+1$ Cardy states in 
$(p,q)$ minimal superconformal field theory 
are not symmetric under the transformation of $(-1)^{f_R}$ 
($f_R$ is the worldsheet fermion number), 
even though the superconformally invariant boundary conditions \eq{scibc}
of $\eta=\pm1$ are exchanged for each other. 
The superstring Cardy state is also not an exception 
but we show that 
one of the branes obtained 
from the transformation of $(-1)^{f_R}$ comes 
to play a role of 
fundamental degrees of freedom, 
principal FZZT branes \cite{SeSh}. 
The principal $\eta=-1/+1$ FZZT branes 
of this theory are simply related 
under the transformation of $(-1)^{f_R}$, 
and actually we show that 
they do not satisfy the Cardy equations among each other. 
This is reminiscent of 
order/disorder parameters of the Kramers-Wannier duality. 
We actually argue that 
they are not mutually local in the superstring spacetime, 
by evaluating annulus amplitudes (section 3). 

The organization of this paper is as follows: 
In section 2, we summarize the Cardy states of this theory. 
After we briefly review the definition of Cardy states in section 2.1, 
we show the way to derive the remaining wave functions 
in section 2.2 and the boundary states of 
our superstring theory are discussed in section 2.3. 
In section 3, we evaluate various annulus amplitudes of FZZT branes. 
Section 4 is devoted to conclusion and discussion.

\section{Boundary states of $(p,q)$ minimal superstring theory}

\subsection{Cardy states in superconformal field theory}
We now recall the definition of 
Cardy states \cite{cardy1} 
in the SCFT case to fix our notation. 
This case (but only odd models) 
was studied under the spin-model GSO projection \cite{SCardy}. 
The following discussion does not require any restrictions, 
and any GSO projections are not imposed form the beginning.

Superconformally invariant boundary states $\ket{B;\eta}$ 
are defined as
\begin{align}
(L_n-\bar L_{-n})\ket{B;\eta}=(G_r-i\eta\, \bar G_{-r})\ket{B;\eta}=0, \qquad 
(\eta=\pm1;\,n\in \mathbb Z,\ r\in \mathbb Z +\nu).\label{SCinv}
\end{align}
Here 
$L_n \,(\bar L_n)$ and $ G_r \,(\bar G_r)$ are 
the left (right) handed super-Virasoro generators that satisfy 
\begin{align}
[L_n,L_m]&=(n-m)L_{n+m} + \frac{\hat c}{8}(n^3-n)\delta_{n+m,0} \nn\\
[L_n,G_r]&=(n-r)G_{n+r} \nn\\
\{G_r,G_s\}&=2L_{r+s}+\frac{\hat c}{2}(r^2-\frac{1}{4})\delta_{r+s,0}, 
\end{align}
and $\nu=1/2$ (or $\nu=0$) when $\ket{B;\eta}$ is in NSNS (or RR) sector. 
The closed string Hilbert space can be 
expanded into irreducible Verma modules of super-Virasoro algebra 
or superconformal families
\begin{align}
\mathcal H^{\rm (c)} =\bigoplus_{i\in NSNS,RR} \mathcal V_i \otimes \bar {\mathcal V_{i}} 
\equiv \bigoplus_{i\in NSNS,RR} \bigl[\phi_i(z,\bar z)\bigr].
\end{align}
When the left and right Verma modules are isomorphic, 
$\mathcal V_i \cong \bar{\mathcal V_i}$, 
one can find the Ishibashi state \cite{Ishibashi} 
in these irreducible Verma modules, 
\begin{align}
\ket{i;\eta}\bigl> =\sum_{N}\ket{i;N}\otimes U_\eta A 
\overline{ \ket{i;N^*}},
\end{align}
which satisfies \eq{SCinv}. 
Here $A$ is an anti-unitary operator that commutes with the above 
super-Virasoro generators, 
$U_\eta$ is an automorphism
of irreducible Verma module $\bar {\mathcal V_i}$,
\begin{align}
U_\eta \bar L_n U_\eta^{-1}=\bar L_n,\qquad 
U_\eta \bar G_r U_\eta^{-1} = i\eta \,\bar G_r (-1)^{f_R}, \qquad 
U_\eta^\dagger=U_\eta^{-1},
\end{align} 
and $\overline{\ket{i;N^*}}$ is a hermitian conjugation 
(defined by $\bar G_r^\dagger =\bar G_{-r}, \bar L_n^\dagger=\bar L_{-n}$ ) 
of a dual base $\overline{\bra{i;N^*}}$,%
\footnote{In the case of unitary CFTs, we can take orthonormal bases; 
Since we have to treat nonunitary CFTs in general, we use dual bases. 
Our construction of these Ishibashi states is
based on Watt's technique noted in \cite{bppz}.}
\begin{align}
\overline{\bracket{i;N^*}{j;M}} = \overline{\bracket{i}{j}}\, \delta_{N,M}.
\end{align}
Note that the chirality $\Gamma$ on each states is given as%
\footnote{We can also define the chirality of the Ramond-Ramond 
ground state as 
$\Gamma\, \ket{\theta_{RR};\eta}\bigr>=-\ket{i_{RR};\eta}\bigr>$. }
\begin{align}
\Gamma\, \ket{i_{NSNS};\eta}\bigr>=+\ket{i_{NSNS};\eta}\bigr>,
\quad 
\Gamma\, \ket{i_{RR};\eta}\bigr>=\eta\,\ket{i_{RR};\eta}\bigr>,
\quad
\Gamma\, \ket{\theta_{RR};\eta}\bigr>=+\ket{i_{RR};\eta}\bigr>,
\end{align}
and states of different $\eta$ are related as  
$\ket{i;-\eta}\bigr>=(-1)^{f_R}\ket{i;\eta}\bigr>$. 
Also note that RR zero-mode representations 
in these Ishibashi states are given as 
\begin{align}
\ket{j_{RR};\eta}\bigr> = \ket{j+}\otimes\overline{\ket{j+}} 
+i\eta\, \ket{j-}\otimes\overline{\ket{j-}} +\cdots,
\end{align}
and $\bracket{j+}{j+}=\bracket{j-}{j-}=1$. 

Annulus amplitudes between these Ishibashi states are then given with 
the moduli parameter $\rmq=\bar \rmq=e^{-2\pi t}$ as 
\begin{align}
\bigl<\bra{i;\eta} \rmq^{\frac{1}{2}(L_0-\frac{\hat c}{16})}
\bar \rmq^{\frac{1}{2}(\bar L_0-\frac{\hat c}{16})} \ket{j;\eta'}\bigr> 
=\mathcal G_{ij}\,\tr_{i}(\eta\eta')^f \rmq^{L_0-\frac{\hat c}{16}}.
\end{align}
Here we denote $\mathcal G_{ij}\equiv \bracket{\phi_i}{\phi_j}$ 
as the Zamolodchikov metric of superconformal primary fields 
$\phi_i(z,\bar z)$. It is convenient to introduce 
dual superconformal primary fields 
$\phi_{i^*}\equiv \mathcal G^{ij}\phi_j$, then we can write the 
formal completeness relation as
\begin{align}
\bunit_\eta
=\sum_{i,j} \ket{i,\eta}\!\bigr> \mathcal G^{ij}\bigl<\!\bra{j,\eta} 
=\sum_{i} \ket{i,\eta}\!\bigr> \bigl<\!\bra{i^*,\eta}. 
\end{align}
Considering this relation, closed-channel amplitudes 
between general boundary states 
$\ket{\alpha;\eta}\equiv \sum_i \ket{i;\eta}\!\bigr>\, 
\langle\!\bracket{i^*;\eta}{\alpha}$ 
are expressed as
\begin{align}
\bra{\alpha;\eta} \rmq^{\frac{1}{2}(L_0-\frac{\hat c}{16})}
\bar \rmq^{\frac{1}{2}(\bar L_0-\frac{\hat c}{16})} \ket{\beta;\eta'}
=\sum_i \bracket{\alpha}{i;\eta}\!\rangle
\langle\!\bracket{i^*;\eta'}{\beta} 
\,\tr_{i}(\eta\eta')^f \rmq^{L_0-\frac{\hat c}{16}}. \label{ClosedCha}
\end{align}

On the other hand, open channel amplitudes are given as a sum over the 
open channel Hilbert space $\mathcal H_{\alpha\beta}^{\rm (o)}$ 
with the boundaries. Therefore they must be expanded 
into a sum of Virasoro characters 
with non-negative integer $n_{\alpha\beta}{}^{h}$ \cite{cardy1}:
\begin{align}
\tr_{\mathcal H_{\alpha\beta}^{\rm (o)}} \tilde \rmq^{L_0-\frac{\hat c}{16}}
=\sum_h n_{\alpha\beta}{}^{h} \tr_h \tilde \rmq^{L_0-\frac{\hat c}{16}}, 
\label{OpenCha}
\end{align}
with $\tilde \rmq=e^{-2\pi/t}$. 
Note that the label $h$ runs 
among irreducible Virasoro primary states 
belonging to $\text{NS}\pm$ or $\text{R}\pm$ sector 
in open channel \cite{SCardy}.
Comparing both expressions \eq{ClosedCha} and \eq{OpenCha}, we obtain 
\begin{align}
0&=\sum_{i\in NSNS,\, RR}\Bigl( 
\bracket{\alpha}{i;\eta}\!\rangle
\langle\!\bracket{i^*;\eta}{\beta} -\sum_{h\in NS\pm} n_{\alpha\beta}{}^{h} 
S_h{}^{i}
\Bigr)
\tr_i \rmq^{L_0-\frac{\hat c}{16}}, \nn\\
0&=\sum_{i\in NSNS,\, RR}\Bigl( 
\bracket{\alpha}{i;\eta}\!\rangle
\langle\!\bracket{i^*;-\eta}{\gamma} -\sum_{h\in R\pm} n_{\alpha\gamma}{}^{h} 
S_h{}^{i}
\Bigr)
\tr_i (-1)^f \rmq^{L_0-\frac{\hat c}{16}}, \label{CardyEq0}
\end{align}
where $\{S_l{}^i\}$ are modular matrices of the characters
under $S$ transformation, $\tau \to -1/\tau$, which we define as, 
\begin{align}
\chi_{h_{NS\pm}}(-1/\tau)
\equiv& \sum_{i}S_{h_{NS\pm}}{}^i \chi_i(\tau) 
= \frac{1}{2}\sum_{i_{NS}} S_{h_{NS}}{}^{i_{NS}}
\,\chi_{i_{NS}}(\tau)
\pm \frac{1}{2\sqrt 2} 
\sum_{i_{R}} S_{h_{\widetilde{NS} }}{}^{i_{R}}
\,\chi_{i_R}(\tau), \nn\\
\chi_{h_{R\pm}}(-1/\tau)
\equiv& \sum_iS_{h_{R\pm}}{}^i\chi_i(\tau) 
=\frac{\sqrt 2}{2}\sum_{i_{\widetilde{NS} }} S_{h_{R}}{}^{i_{\widetilde{NS} }}
\,\chi_{i_{\widetilde{NS}}}(\tau),  \nn\\
\chi_{\theta_{R\pm}}(-1/\tau)
\equiv& \sum_iS_{\theta_{R\pm}}{}^i\chi_i(\tau) 
=\frac{1}{2\sqrt 2}\sum_{i_{\widetilde{NS} }} S_{h_{R}}{}^{i_{\widetilde{NS} }}
\,\chi_{i_{\widetilde{NS}}}(\tau) 
\pm \frac{1}{2} 
\delta_{h_R,\theta_R}.
\label{Strans}
\end{align}
Noting that the characters in $\widetilde{\rm R}$ sector must vanish due to 
fermion zero modes, except for the Ramond ground states 
(here we define $\chi_{\theta_R}^{\widetilde R}(\tau)=1$), 
we obtain the following Cardy equations: 
\begin{align}
0&=
\bracket{\alpha}{i;\eta}\!\rangle
\langle\!\bracket{i^*;\eta}{\beta} -\sum_{h\in NS\pm} n_{\alpha\beta}{}^{h} 
S_h{}^{i}
, \quad (i\in NSNS,RR), \nn\\
0&=
\bracket{\alpha}{i;\eta}\!\rangle
\langle\!\bracket{i^*;-\eta}{\gamma} -\sum_{h\in R\pm} n_{\alpha\gamma}{}^{h} 
S_h{}^{i}, \quad (i\in NSNS), \nn\\
0&=
\bracket{\alpha}{\theta_{RR};\eta}\!\rangle
\langle\!\bracket{\theta_{RR}^*;-\eta}{\gamma} 
-\sum_{h=\theta_{R\pm}} n_{\alpha\gamma}{}^{\theta_R}.
\label{CardyEq1}
\end{align}
Then, following the usual procedure 
(identifying $n_{\alpha\beta}{}^{h}$ with the 
fusion number $\mathcal N_{\alpha\beta}{}^{h}$ \cite{cardy1} 
under the Verlinde formula \cite{verlinde}
as a non-negative integer valued 
matrix representation of fusion algebra \cite{bppz} and considering the 
trivial relation $\mathcal N_{0_{{\rm NS}+} \beta}{}^h=\delta_{\beta}{}^{h}$), 
we can define the Cardy states $\ket{h_{NS\pm}}$ and $\ket{h_{R\pm}}$ 
for each Virasoro highest weight $h_{NS_\pm}$ and $h_{R_\pm}$ in open channel:
\begin{align}
\ket{h_{NS\pm}} 
 &=\frac{1}{\sqrt{2}}\sum_{i\in NS}\frac{S_{h_{NS}}{}^{i_{NS}}}
    {\sqrt{S_{0_{NS}}{}^{i_{NS}}}} 
     \ket{i_{NSNS};\eta=-1}\bigr>
  \pm\frac{1}{2^{3/4}}\sum_{i\in R} \frac{S_{h_{\widetilde {NS}}}{}^{i_R}}
{\sqrt{S_{0_{\widetilde{NS}}}{}^{i_R}}} 
\ket{i_{RR};\eta=-1}\bigr>, \nn\\
\ket{h_{R\pm}}
 &=\sum_{i\in NS}\frac{S_{h_R}{}^{i_{\widetilde{NS}}}}
    {\sqrt{S_{0_{NS}}{}^{i_{NS}}}} 
     \ket{i_{NSNS};\eta=+1}\bigr>
  \pm\frac{\sqrt{2}}{2^{3/4}}\sum_{i\in R} 
    \frac{\psi_{h_{\widetilde R}}{}^{i_{\widetilde R}}}
{\sqrt{S_{0_{\widetilde {NS}}}{}^{i_{R}}}} 
\ket{i_{RR};\eta=+1}\bigr>, \nn\\
\ket{\theta_{R\pm}}
 &=\frac{1}{2}\sum_{i\in NS}\frac{S_{\theta_R}{}^{i_{\widetilde{NS}}}}
    {\sqrt{S_{0_{NS}}{}^{i_{NS}}}} 
     \ket{i_{NSNS};\eta=+1}\bigr>
  \pm\frac{1}{2^{1/4}}\sum_{i\in R} 
    \frac{\psi_{h_{\widetilde R}}{}^{i_{\widetilde R}}}
{\sqrt{S_{0_{\widetilde {NS}}}{}^{i_{R}}}} 
\ket{i_{RR};\eta=+1}\bigr>, \label{generalCardy}
\end{align}
with 
$\psi_{\theta_{\widetilde R}}{}^{\theta_{\widetilde R}}
=1\ (=S_{\theta_{\widetilde R}}{}^{\theta_{\widetilde R}})$ 
for the Cardy state of the Ramond ground state, $\ket{\theta_{R_\pm}}$. 
Here we denote the identity operator as ``0'', 
and $\sqrt{S_{0_{NS}}{}^{i_{NS}}}$ is formally defined as it satisfies 
\begin{align}
\sqrt{S_{0_{NS}}{}^{i_{NS}}} 
\cdot \sqrt{S_{0_{NS}}{}^{i^*_{NS}}}
\equiv S_{0_{NS}}{}^{i_{NS}}, \quad
\sqrt{S_{0_{\widetilde{NS}}}{}^{i_{R}}} 
\cdot \sqrt{S_{0_{\widetilde{NS}}}{}^{i^*_{R}}}
\equiv S_{0_{\widetilde{NS}}}{}^{i_{R}}.
\end{align}
The normalization of $\psi$ is chosen as above 
for later convenience. 

Here we consider the spin-model GSO projection 
($\Gamma=-1$ on RR sector) \cite{fms} for odd models. 
In this case, 
the last wave functions $\psi_h{}^i$ of $\ket{h_\pm}$ 
can be consistently dropped since all the characters in 
${\widetilde R}$ sector vanish (see also the Verlinde 
formula of this case \cite{geneVerlinde}). 
This gives the previous result \cite{SCardy}:%
\footnote{Note that our normalization of RR Ishibashi states is 
differed by $\sqrt 2$ from \cite{SCardy}, that is 
$\ket{i_{RR};\eta=\pm1}\bigr>_{here}
=\sqrt 2 \ket{i_{RR};\eta=\pm1}\bigr>_{there}$} 
\begin{align}
\ket{h_{NS\pm}} 
 &=\frac{1}{\sqrt{2}}\sum_{i\in NS}\frac{S_{h_{NS}}{}^{i_{NS}}}
    {\sqrt{S_{0_{NS}}{}^{i_{NS}}}} 
     \ket{i_{NSNS};\eta=-1}\bigr>
  \pm\frac{1}{2^{3/4}}\sum_{i\in R} \frac{S_{h_{\widetilde {NS}}}{}^{i_R}}
{\sqrt{S_{0_{\widetilde{NS}}}{}^{i_R}}} 
\ket{i_{RR};\eta=-1}\bigr> \nn\\
\ket{h_{R+}}
 &=\sum_{i\in NS}\frac{S_{h_R}{}^{i_{\widetilde{NS}}}}
    {\sqrt{S_{0_{NS}}{}^{i_{NS}}}} 
     \ket{i_{NSNS};\eta=+1}\bigr>. 
\end{align}
For even models, since we should be careful about 
the Ramond ground state $\ket{\theta_{R\pm}}$
and need to know the remaining wave functions $\psi_h{}^{i}$,  
we will discuss this case in next subsection.

\subsection{The wave functions $\psi$ in each SCFT}

In this subsection, 
we determine the remaining wave functions $\psi$ 
in each superconformal field theories. 
As we noted in section 1, we consider the following fusion rule:
\begin{align}
n_{(1,2)_+,(k,l)_+}{}^{(r,s)}
=2\,\mathcal N_{(1,2)_+,(k,l)_+}{}^{(r,s)}
= 2\,\delta_{(k,l+1)+}{}^{(r,s)}+2\,\delta_{(k,l-1)_-}{}^{(r,s)}. 
\end{align}
of the simplest degenerate primary $(1,2)_+$ of Virasoro algebra \cite{bpz}. 
The factor 2 in front of $\mathcal N_{i,j}{}^k$ comes from the degeneracy of 
Ramond zero-mode representations.%
\footnote{See e.g., \cite{geneVerlinde}. This naturally comes from the 
Verlinde formula of super-Virasoro algebra.}
We first reconsider the case of 
super Liouville field theory and see that this procedure 
actually reproduces the results of \cite{fuku-hoso}. 
We then discuss the case of $(p,q)$ minimal superconformal field theory. 
We also summarize the corresponding modular matrices in each SCFT. 

\subsubsection{the super-Liouville case}

The corresponding Cardy states were 
found in \cite{fuku-hoso,ars} and further discussed in 
\cite{newhat,UniCom,SeSh,SeSh2}.%
\footnote{Note that the case of $\mathcal N=2$ 
super-Liouville theory was investigated in 
\cite{EguchiSugawara}}
They are expanded in the off-shell Hilbert space \cite{Seiberg},
\begin{align}
\mathcal H^{({\rm c})}=
\bigotimes_{\nu_{\rm NSNS,RR}>0} \, 
\mathcal V_\nu \otimes \widetilde{\mathcal V_\nu} 
=\bigotimes_{\nu_{\rm NSNS,RR}>0}\, 
[e^{(Q/2+i\nu)\phi(z,\bar z)}],
\end{align}
and the Cardy states 
for non-degenerate representations, $\Delta_\sigma=Q^2/8+\sigma^2/2$, are 
\begin{align}
\ket{\sigma_{NS\pm}}&=\int_0^\infty d\nu \,\biggl( \ 
\frac{1}{\sqrt 2}\bigl( 2\cosh (2\pi i \sigma \nu)\bigr) A^{(L)}_{NS}(\nu) 
    \ket{\nu_{NSNS};\eta=-1}\bigr>\,\pm \nn\\
&\qquad\qquad\ \pm 
\frac{1}{2^{3/4}}\bigl(2 \cosh (2\pi i \sigma \nu )\bigr) A^{(L)}_{R}(\nu) 
       \ket{\nu_{RR};\eta=-1}\bigr>
\biggr), \nn\\
\ket{\sigma_{R\pm}}&=\int_0^{\infty} d\nu \,\biggl(  
\bigl(2 \cosh (2\pi i \sigma \nu)\bigr) A^{(L)}_{NS}(\nu)
    \ket{\nu_{NSNS};\eta=+1}\bigr>\,\pm \nn\\
&\qquad\qquad\  \pm 
\frac{\sqrt 2}{2^{3/4}}\bigl(2i\sinh (2\pi i \sigma \nu )\bigr) A^{(L)}_{R}(\nu)
       \ket{\nu_{RR};\eta=+1}\bigr>
\biggr),
\end{align}
and those for degenerate representations $\sigma=i(nb+m/b)$ are 
\begin{align}
\ket{(n,m)_{NS\pm}}&=\int_0^\infty d\nu \,\biggl( \ 
\frac{1}{\sqrt 2}\Bigl(4 \sinh \bigl(\pi nb\nu\bigr)
                  \sinh \bigl(\frac{\pi m\nu}{b}\bigr) \Bigr)
                  A^{(L)}_{NS}(\nu) 
                  \ket{\nu_{NSNS};\eta=-1}\bigr>
                  \,\pm \nn\\
&\qquad\qquad\ \pm 
\frac{1}{2^{3/4}} \Bigl(4 \sinh \bigl(\pi nb\nu+\frac{i\pi n}{2}\bigr)
                  \sinh \bigl(\frac{\pi m\nu}{b}-\frac{i\pi n}{2}\bigr) \Bigr)
                  A^{(L)}_{R}(\nu) 
                  \ket{\nu_{RR};\eta=-1}\bigr>
\biggr), \nn\\
\ket{(n,m)_{R\pm}}&=\int_0^{\infty} d\nu \,\biggl(  
               \Bigl( 4 \sinh \bigl(\pi nb\nu \bigr) 
                  \sinh \bigl(\frac{\pi m\nu}{b}\bigr) \Bigr)
                  A^{(L)}_{NS}(\nu)
                  \ket{\nu_{NSNS};\eta=+1}\bigr>
                  \,\pm \nn\\
&\qquad\qquad\  \pm 
\frac{\sqrt 2}{2^{3/4}}
                \Bigl(4i  \sinh \bigl(\pi nb \nu +\frac{i\pi n }{2}\bigr) 
                  \cosh \bigl(\frac{\pi m\nu}{b}-\frac{i\pi n}{2}\bigr) \Bigr)
                  A^{(L)}_{R}(\nu)
                  \ket{\nu_{RR};\eta=+1}\bigr>
\biggr),
\end{align}
where $A^{(L)}_{NS}(\nu)$ and $A^{(L)}_R(\nu)$ are defined as 
\begin{align}
A^{(L)}_{NS}(\nu)&
  \equiv 
     \frac{\Gamma(1-i\nu b)\Gamma(1-i\nu/b)}{2\pi \nu}\, \mu^{-i\nu/b} 
       =1/\sqrt{S_{0_{NS}}{}^{\nu_{NS}}},  \nn\\
A^{(L)}_{R}(\nu)&
  \equiv
     \frac{\Gamma(1/2-i\nu b)\Gamma(1/2-i\nu/b)}{2\pi}\, \mu^{-i\nu/b}
       =1/\sqrt{S_{0_{\widetilde{NS}}}{}^{\nu_{R}}}.
\end{align}
The modular bootstrap method, 
except for the $\widetilde{\rm R}$ wave function $\psi$, 
was studied in \cite{fuku-hoso,ars}. 
The corresponding modular matrices are actually obtained as 
\begin{align}
S^{(L)}_{\sigma_{NS}}{}^{\nu_{NS}}=S^{(L)}_{\sigma_{\widetilde{NS}}}{}^{\nu_R}
=S^{(L)}_{\sigma_{R}}{}^{\nu_{\widetilde{NS}}}
&=2\cosh \bigl(2\pi i \sigma\nu\bigr) \nn\\
S^{(L)}_{(n,m)_{NS}}{}^{\nu_{NS}}=
S^{(L)}_{(n,m)_R}{}^{\nu_{\widetilde{NS}}}
                &=4\sinh \bigl(\pi nb\nu\bigr)
                  \sinh \bigl(\frac{\pi m\nu}{b}\bigr) \nn\\
S^{(L)}_{(n,m)_{\widetilde{NS}}}{}^{\nu_R}&=
                 4\sinh \bigl(\pi nb\nu+\frac{i\pi n}{2}\bigr)
                  \sinh \bigl(\frac{\pi m\nu}{b}-\frac{i\pi n}{2}\bigr).
\end{align}
form the characters:
\begin{align}
\chi_\sigma^{(NS)}(\tau)=\rmq^{\frac{\sigma^2}{2}}\chi_0^{(NS)}(\tau),\quad
\chi_\sigma^{(\widetilde{NS})}(\tau)=\rmq^{\frac{\sigma^2}{2}} 
  \chi_0^{(\widetilde{NS})}(\tau),\quad
\chi_\sigma^{(R)}(\tau)=\rmq^{\frac{\sigma^2}{2}}\chi_0^{(R)}(\tau)
\end{align}
for non-degenerate representations, and 
\begin{align}
\chi_{(n,m)}^{(NS)}(\tau) 
&\equiv \chi_{\frac{i}{2}(nb+m/b)}^{(NS)}(\tau)
-\chi_{\frac{i}{2}(nb-m/b)}^{(NS)}(\tau), \nn\\
\chi_{(n,m)}^{(\widetilde{NS})}(\tau)
&\equiv \chi_{\frac{i}{2}(nb+m/b)}^{(\widetilde{NS})}(\tau)
-(-1)^{mn} \chi_{\frac{i}{2}(nb-m/b)}^{(\widetilde{NS})}(\tau),\nn\\
\chi_{(n,m)}^{(R)}(\tau) 
&\equiv \chi_{\frac{i}{2}(nb+m/b)}^{(R)}(\tau)
-\chi_{\frac{i}{2}(nb-m/b)}^{(R)}(\tau)
\end{align}
for degenerate representations. 
See appendix \ref{modfunc} for our definition of basic modular 
functions (like $\chi_0^{(NS)}(\tau)$) and its modular transformation. 

The remaining wave functions $\psi$ were then obtained from the 
conformal bootstrap method \cite{fuku-hoso}. 
Here we show that they can also be obtained 
from the modular bootstrap method 
by considering the simplest degenerate primary 
operator $(1,2)_+$ and its fusion rule $\mathcal N_{i,j}{}^k$ 
with operators in ${\rm R_+}$ sector. 
Actually, the fusion rule between $(1,2)_+$ and 
non-degenerate primary ${\sigma_{\rm R+}}$ 
is controllable \cite{doZZ,tes1} and leads to
\begin{align}
n_{(1,2)_+,(n,m)_+}{}^h=2\mathcal N_{(1,2)_+,(n,m)_+}{}^h
= 2\delta_{(n,m+1)+}{}^h+2\delta_{(n,m-1)_-}{}^h, \nn\\
n_{(1,2)_+,\sigma_{R+}}{}^h=2\mathcal N_{(1,2)_+,\sigma_{R+}}{}^h
= 2\delta_{(\sigma+ib/2)_{R+}}{}^h+2\delta_{(\sigma-ib/2)_{R-}}{}^h.
\end{align}
Note that the chirality of the second terms is flipped due to the 
boundary Liouville action (see e.g., \cite{fuku-hoso,ars}) 
whose fermion number is odd, 
and that the factor ``2'' follows from 
the wave functions of NSNS sector. 
The corresponding Cardy equations are 
\begin{align}
&\bracket{(1,2)_+}{\nu_{RR};\eta=+1}\!\rangle
\langle\!\bracket{-\nu_{RR};\eta=+1}{(1,2)_+} \nn\\
&\qquad =\frac{1}{\sqrt 2}\bigl(
 +S_{(1,3)_{\widetilde{\rm NS}}}{}^{\nu_{\widetilde{\rm R}}}
 -S_{(1,1)_{\widetilde{\rm NS}}}{}^{\nu_{\widetilde{\rm R}}}
 \bigr) \nn\\
&\qquad =\frac{1}{\sqrt{2}}
\frac{\bigl(4i\cosh(\pi b\nu)\sinh(2\pi \nu/b)\bigr)\cdot
      \bigl(4i\cosh(-\pi b\nu)\sinh(-2\pi \nu/b)\bigr)}
{4\cosh(\pi b\nu)\cosh(\pi\nu/b)} \nn\\
&\bracket{(n,m)_+}{\nu_{RR};\eta=+1}\!\rangle
\langle\!\bracket{-\nu_{RR};\eta=+1}{(1,2)_+} \nn\\
&\qquad =\frac{1}{\sqrt 2}\bigl(
 +S_{(n,m+1)_{\widetilde{\rm NS}}}{}^{\nu_{\widetilde{\rm R}}}
 -S_{(n,m-1)_{\widetilde{\rm NS}}}{}^{\nu_{\widetilde{\rm R}}}
 \bigr) \nn\\
&\qquad =\frac{1}{\sqrt{2}}
\frac{\bigl(4i\sinh(n\pi b\nu+\frac{\pi ni}{2})
       \sinh(m\pi \nu/b-\frac{\pi ni}{2})\bigr)\cdot
      \bigl(4i\cosh(-\pi b\nu)\sinh(-2\pi \nu/b)\bigr)
}
{4\cosh(\pi b\nu)\cosh(\pi\nu/b)}\nn\\
&\bracket{\sigma_{R+}}{\nu_{RR};\eta=+1}\!\rangle
\langle\!\bracket{-\nu_{RR};\eta=+1}{(1,2)_+} \nn\\
&\qquad =\frac{1}{\sqrt 2}\bigl(
 +S_{(\sigma+ib/2)_{\widetilde{\rm NS}}}{}^{\nu_{\widetilde{\rm R}}}
 -S_{(\sigma-ib/2)_{\widetilde{\rm NS}}}{}^{\nu_{\widetilde{\rm R}}}
 \bigr) \nn\\
&\qquad =\frac{1}{\sqrt{2}}
\frac{\bigl(2i\sinh(2\pi i\sigma\nu)\bigr)\cdot 
      \bigl(4i\cosh(-\pi b\nu)\sinh(-2\pi \nu/b)\bigr)
}
{4\cosh(\pi b\nu)\cosh(\pi\nu/b)},
\end{align}
and solving these equation we can obtain 
\begin{align}
\psi_{\sigma_{\widetilde{R}}}{}^{\nu_{\widetilde{R}}}
&=2i\sinh \bigl(2\pi i \sigma \nu\bigr), \nn\\
\psi_{(n,m)_{\widetilde{R}}}{}^{\nu_{\widetilde{R}}}  &=
                 4i\sinh \bigl(\pi nb\nu +\frac{i\pi n}{2}\bigr) 
                  \cosh \bigl(\frac{\pi m\nu}{b} -\frac{i\pi n}{2}\bigr),
\end{align}
and this correctly reproduces the results of \cite{fuku-hoso}.

\subsubsection{the $(p,q)$ minimal superconformal field theory case}
Next we apply the above argument to the case of $(p,q)$ minimal 
superconformal field theory. 
This theory can be classified into two categories: 
even and odd models \cite{pqsmini}. 
The modular matrices among NS, ${\rm \widetilde {NS}}$ and R sector 
were derived in \cite{modSsmini} from the character formula \cite{gko},
\begin{align}
\chi_{(r,s)}^{(NS)}(\tau)
&=\chi_0^{(NS)}(\tau)\sum_{n\in \mathbb Z} 
\Bigl[
\rmq^{\frac{(2npq+qr-ps)^2}{8pq}} - \rmq^{\frac{(2npq+qr+ps)^2}{8pq}}
\Bigr], \nn\\
\chi_{(r,s)}^{(\widetilde{NS})}(\tau)
&=\chi_0^{(\widetilde{NS})}(\tau)\sum_{n\in \mathbb Z} (-1)^{2pq}
\Bigl[
\rmq^{\frac{(2npq+qr-ps)^2}{8pq}} - (-1)^{rs} \rmq^{\frac{(2npq+qr+ps)^2}{8pq}}
\Bigr], \nn\\
\chi_{(r,s)}^{(R)}(\tau)
&=\chi_0^{(R)}(\tau)\sum_{n\in \mathbb Z} 
\Bigl[
\rmq^{\frac{(2npq+qr-ps)^2}{8pq}} - \rmq^{\frac{(2npq+qr+ps)^2}{8pq}}
\Bigr], \nn\\
\chi_{(\frac{p}{2},\frac{q}{2})}^{(R)}(\tau)
&=\frac{1}{2}\chi_0^{(R)}(\tau)\sum_{n\in \mathbb Z} 
\Bigl[
\rmq^{\frac{(2npq)^2}{8pq}} - \rmq^{\frac{(2npq+pq)^2}{8pq}}
\Bigr]\quad (\hbox{only even model}),
\end{align}
as 
\begin{align}
S_{(r,s)}{}^{(\bar r,\bar s)} = \frac{4}{\sqrt{pq}} 
(-1)^{\frac{(r-s)(\bar r-\bar s)}{2}} 
\sin \bigl(\frac{r\bar r}{2p}(q-p)\pi\bigr)
\sin \bigl(\frac{s\bar s}{2q}(q-p)\pi\bigr).
\end{align}

For the ${\rm \widetilde R}$ wave function 
$\psi_{(r,s)}{}^{(\bar r,\bar s)}$, 
we should know the fusion rule of open strings. 
One way to know is the super Coulomb gas formalism 
in Ramond sector \cite{corsmini}. 
Since the chirality of the screening charges is odd, 
The fusion rule is also given in the following form:
\begin{align}
n_{(1,2)_+,(k,l)_+}{}^{(r,s)}
=2\,\mathcal N_{(1,2)_+,(k,l)_+}{}^{(r,s)}
= 2\,\delta_{(k,l+1)+}{}^{(r,s)}+2\,\delta_{(k,l-1)_-}{}^{(r,s)}. 
\label{fusionmm}
\end{align}
This leads to the Cardy equations,
\begin{align}
&\bracket{(1,2)_+}{(r,s)_{RR};\eta=+1}\!\rangle
\langle\!\bracket{(-r,-s)_{RR};\eta=+1}{(1,2)_+} \nn\\
&\qquad =\frac{1}{\sqrt 2}\bigl(
 +S_{(1,3)_{\widetilde{\rm NS}}}{}^{(r,s)_{{\rm R}}}
 -S_{(1,1)_{\widetilde{\rm NS}}}{}^{(r,s)_{{\rm R}}}
 \bigr) \nn\\
&\qquad =\frac{-1}{\sqrt{2}}
  \dfrac{
     \Bigl(\dfrac{4}{\sqrt{pq}}
       \sin\bigl(\dfrac{r}{2p}(q-p)\pi\bigr)
       \sin\bigl(\dfrac{2s}{2q}(q-p)\pi\bigr)
     \Bigr)^2
  }{
      \dfrac{4}{\sqrt{pq}}
       \sin\bigl(\dfrac{r}{2p}(q-p)\pi\bigr)
       \sin\bigl(\dfrac{s}{2q}(q-p)\pi\bigr)
    }, \nn\\
&\bracket{(k,l)_+}{(r,s)_{RR};\eta=+1}\!\rangle
\langle\!\bracket{(-r,-s)_{RR};\eta=+1}{(1,2)_+} \nn\\
&\qquad =\frac{1}{\sqrt 2}\bigl(
 +S_{(k,l+1)_{\widetilde{\rm NS}}}{}^{(r,s)_{{\rm R}}}
 -S_{(k,l-1)_{\widetilde{\rm NS}}}{}^{(r,s)_{{\rm R}}}
 \bigr) \nn\\
&\qquad =\frac{1}{\sqrt{2}} (-1)^{\frac{(k-l-1)(r-s)}{2}}\times \nn\\
&\quad \times 
  \frac{
    \Bigl(\dfrac{4}{\sqrt{pq}}
       \sin\bigl(\dfrac{kr}{2p}(q-p)\pi\bigr)
       \sin\bigl(\dfrac{ls}{2q}(q-p)\pi\bigr)
     \Bigr)\cdot
    \Bigl(\dfrac{4}{\sqrt{pq}}
       \sin\bigl(\dfrac{r}{2p}(q-p)\pi\bigr)
       \sin\bigl(\dfrac{2s}{2q}(q-p)\pi\bigr)
     \Bigr)
  }{
     \dfrac{4}{\sqrt{pq}}
       \sin\bigl(\dfrac{r}{2p}(q-p)\pi\bigr)
       \sin\bigl(\dfrac{s}{2q}(q-p)\pi\bigr)
   }.
\end{align}
Solving these equations, we obtain 
\begin{align}
\psi_{(k,l)}{}^{(r,s)}=\frac{4}{\sqrt{pq}}
    (-1)^{\frac{(k-l)(r-s)+1}{2}}
    \sin\bigl(\dfrac{kr}{2p}(q-p)\pi\bigr)
    \sin\bigl(\dfrac{ls}{2q}(q-p)\pi\bigr). \label{matwf}
\end{align}
From this formula, 
we can see that any $(k,l)\neq (p/2,q/2)$ Cardy states have 
no contribution from the closed string $(r,s)=(p/2,q/2)$ state. 
From this consideration, we also conclude 
\begin{align}
\psi_{\theta_R}{}^{i_R}= \delta_{\theta_R}{}^{i_R} 
=S_{\theta_{\widetilde R}}{}^{i_{\widetilde R}}. 
\end{align}

For an exercise of the later discussion, we also consider 
the spin-model GSO projections \cite{fms} 
for even models. 
Here we first assume 
$\Gamma \ket{\theta_{RR};\eta}\bigr>=-\ket{\theta_{RR};\eta}\bigr>$. 
We then do not have to reconsider the NS Cardy states $\ket{h_{NS\pm}}$ 
and we obtain 
\begin{align}
\ket{h_{NS\pm}} 
 &=\frac{1}{\sqrt{2}}\sum_{i\in NS}\frac{S_{h_{NS}}{}^{i_{NS}}}
    {\sqrt{S_{0_{NS}}{}^{i_{NS}}}} 
     \ket{i_{NSNS};\eta=-1}\bigr>
  \pm\frac{1}{2^{3/4}}\sum_{i\in R} \frac{S_{h_{\widetilde {NS}}}{}^{i_R}}
{\sqrt{S_{0_{\widetilde{NS}}}{}^{i_R}}} 
\ket{i_{RR};\eta=-1}\bigr>, \nn\\
\ket{h_{R+}}
 &=\sum_{i\in NS}\frac{S_{h_R}{}^{i_{\widetilde{NS}}}}
    {\sqrt{S_{0_{NS}}{}^{i_{NS}}}} 
     \ket{i_{NSNS};\eta=+1}\bigr>, \nn\\
\ket{\theta_{R\pm}}
 &=\frac{1}{2}\sum_{i\in NS}\frac{S_{\theta_R}{}^{i_{\widetilde{NS}}}}
    {\sqrt{S_{0_{NS}}{}^{i_{NS}}}} 
     \ket{i_{NSNS};\eta=+1}\bigr>
  \pm\frac{1}{2^{1/4}}
    \frac{1}
{\sqrt{S_{0_{\widetilde {NS}}}{}^{i_{R}}}} 
\ket{\theta_{RR};\eta=+1}\bigr>.
\end{align}
For the case of 
$\Gamma \ket{\theta_{RR};\eta}\bigr>=+\ket{\theta_{RR};\eta}\bigr>$, 
the NS Cardy states $\ket{h_{NS\pm}}$ and $\ket{\theta_{R\pm}}$ 
are no longer Cardy states under the GSO projection. 
We should instead consider the following Cardy states: 
\begin{align}
\ket{h_{NS}} 
&\equiv 
\frac{1}{\sqrt 2}\bigl(\ket{h_{NS+}}+\ket{h_{NS-}}\bigr)
 =\sum_{i\in NS}\frac{S_{h_{NS}}{}^{i_{NS}}}
    {\sqrt{S_{0_{NS}}{}^{i_{NS}}}} 
     \ket{i_{NSNS};\eta=-1}\bigr>, \nn\\
\ket{h_{R}} 
&\equiv \frac{1}{\sqrt 2 }\ket{h_{R\pm}}
 =\frac{1}{\sqrt 2}\sum_{i\in NS}\frac{S_{h_R}{}^{i_{\widetilde{NS}}}}
    {\sqrt{S_{0_{NS}}{}^{i_{NS}}}} 
     \ket{i_{NSNS};\eta=+1}\bigr>, \nn\\
\ket{\theta_{R}} 
&\equiv 
\frac{1}{\sqrt 2}\bigl(\ket{\theta_{R+}}+\ket{\theta_{R-}}\bigr)
 =\frac{1}{\sqrt 2}\sum_{i\in NS}\frac{S_{\theta_R}{}^{i_{\widetilde{NS}}}}
    {\sqrt{S_{0_{NS}}{}^{i_{NS}}}} 
     \ket{i_{NSNS};\eta=+1}\bigr>.
\end{align}
Note that the corresponding fusion rule in this case is 
that of generalized Verlinde formula 
\cite{geneVerlinde} (the fusion rule of super-Virasoro algebra) 
and the states propageting in the open channel are among super-Virasoro 
Verma module (not only among the Virasoro sub-Verma module). 
From this point of view, the factor $1/\sqrt{2}$ 
in the definition of the boundary states is necessary. 
It will be seen that 
the former case is like the type 0B GSO projection and 
the later case is like the type 0A GSO projection in 
minimal superstring theory. 
Although we can also consider other possibility 
(e.g., boundary states with $\eta\to -\eta$), we will not investigate 
this direction here.  

\subsection{Boundary states in $(p,q)$ minimal superstring theory}

We now combine the above Cardy states. 
Since the boundary states in $(p,q)$ minimal superstring theory 
have been discussed in \cite{UniCom,SeSh,SeSh2}, 
we first of all summarize the important things given 
in the previous discussions. 
\begin{itemize}
\item[1.] Since the ghost Ishibashi states in RR sector are 
in $(-1/2,-3/2)$ picture%
\footnote{For a summary of ghost Ishibashi/Cardy states, 
see appendix \ref{ICbg}.}
and $(-1)^{f}=(-1)^{f_L+f_R}=(-1)^{f^L+f^M+1}$, 
we use the following type 0 GSO projection: $(-1)^{f^L+f^M}=+1$ ($-1$) for 
type 0B (type 0A) \cite{UniCom,SeSh}.  
\item[2.] Since 
$(-1)^{f^M} \ket{\theta_{RR};\eta}\bigr>=+\ket{\theta_{RR};\eta}\bigr>$, 
the closed-string contributions of Ramond ground states are denied 
in the boundary states of $\eta=+1$ ($\eta=-1$) branes in the type 0B 
(type 0A) cases \cite{UniCom,SeSh}. 
\item[3.] When we consider the case of negative $\mu<0$, 
we should use the transformation $\mu\to -\mu$ and $\eta^{(L)} \to -\eta^{(L)}$
\cite{SeSh,SeSh2}. That is, 
the boundary states of $\mu<0$ are 
obtained with the following replacement 
in the wave functions of RR emissions:
\begin{align}
&S^{(L)}_{\sigma_{\rm \widetilde {NS}}}{}^{\nu_{\rm R}}
=\bigl(2\cosh(2\pi i\nu\sigma)\bigr)\quad 
\to \quad
\bigl(2\cosh(2\pi i\nu\sigma+\frac{\epsilon\pi i}{2})\bigr),\nn\\
&\psi^{(L)}_{\sigma_{\rm \widetilde {R}}}{}^{\nu_{\rm \widetilde R}}
=\bigl(2i\sinh(2\pi i\nu\sigma)\bigr)\quad 
\to \quad
\bigl(2i\sinh(2\pi i\nu\sigma + \frac{\epsilon\pi i}{2})\bigr), 
\end{align}
where $\epsilon=(1-\sgn(\mu))/2$. 
The corresponding 
boundary cosmological constants are chosen as follows:
\begin{align}
&\zeta= 
\left\{
\begin{array}{ll}
  \sqrt{|\mu|}\cosh(\pi b\sigma) & (\hat \eta=-1) \\
  \sqrt{|\mu|}\sinh(\pi b\sigma) & (\hat \eta=+1)
\end{array}\right.
\end{align}
with the parameter $\hat \eta=\eta\,\sgn(\mu)$ \cite{SeSh}. 
\end{itemize}
Considering this, the boundary states for type 0B theory are 
given as:%
\footnote{Note that the Ramond Cardy state is given as 
$\ket{\sigma,(p/2,q/2)_{\rm R}}=\ket{\sigma,(p/2,q/2)_{\rm R_\pm}}$. } 

\noindent\underline{\bf Type 0B}
\begin{align}
&\ket{\sigma,(k,l)_{\rm NS\pm};\eta=-1}= \nn\\
&\quad=\frac{1}{\sqrt 2}\int_0^\infty d\nu \, \sum_{(r,s)\in \rm NSNS} 
\bigl(2\cosh(2\pi i\nu\sigma)\bigr)\,
S^{(M)}_{(k,l)}{}^{(r,s)} 
A_{NS}(\nu,(r,s)) \ket{\nu,(r,s)_{\rm NSNS};\eta=-1}\bigr> \pm \nn\\
&\quad \qquad \pm \frac{1}{2}\int_0^\infty d\nu \, \sum_{(r,s)\in \rm RR} 
\bigl(2\cosh(2\pi i\nu\sigma+\frac{\epsilon\pi i}{2})\bigr)\,
S^{(M)}_{(k,l)}{}^{(r,s)}
 A_{R}(\nu,(r,s)) \ket{\nu,(r,s)_{\rm RR};\eta=-1}\bigr>, \nn\\
& \hspace{4cm} 
(k+l\in 2\mathbb Z), \label{FZZTem1} \\
&\ket{\sigma,(k,l)_{\rm R\pm};\eta=+1}=\nn\\
&\quad=\frac{1}{\sqrt 2}\int_0^\infty d\nu \, \sum_{(r,s)\in \rm NSNS} 
\bigl(2\cosh(2\pi i\nu\sigma)\bigr)\,
S^{(M)}_{(k,l)}{}^{(r,s)} 
A_{NS}(\nu,(r,s)) \ket{\nu,(r,s)_{\rm NSNS};\eta=+1}\bigr> \pm \nn\\
&\quad \qquad \pm \frac{1}{2}\int_0^\infty d\nu \, \sum_{(r,s)\in \rm RR} 
\bigl(2i\sinh(2\pi i\nu\sigma+\frac{\epsilon\pi i}{2})\bigr)\,
\psi^{(M)}_{(k,l)}{}^{(r,s)}
 A_{R}(\nu,(r,s)) \ket{\nu,(r,s)_{\rm RR};\eta=+1}\bigr>, \nn\\
& \hspace{4cm} 
(k+l\in 2\mathbb Z+1), \label{FZZTep1} 
\end{align}
Here $(r,s)\in {\rm NSNS,\,RR}$ means that $(r,s)$ runs among 
\begin{align}
{\rm NSNS}:&\quad 
1\leq r\leq p-1,\qquad 1\leq s\leq q-1,\qquad qr-ps>0,
\qquad r+s\in 2\mathbb Z,\nn\\
{\rm RR}:&\quad 
1\leq r\leq p-1,\qquad 1\leq s\leq q-1,\qquad qr-ps\ge 0,
\qquad r+s\in 2\mathbb Z+1.
\end{align}
We define 
$A_{\rm X}(\nu,(r,s))\equiv A_{\rm X}^{(L)}(\nu)\cdot A^{(M)}(r,s)$ 
($\rm X= NS \ or\ R$) and  
\begin{align}
A^{(M)}(r,s)\equiv 1/\sqrt{S^{(M)}_{(1,1)}{}^{(r,s)}}, \quad 
A^{(M)}(r,s)\cdot A^{(M)}(-r,-s)=1/{S^{(M)}_{(1,1)}{}^{(r,s)}}.
\end{align}
The Ishibashi states are 
\begin{align}
\ket{\nu,(r,s)_{\rm XX};\eta}\bigr>=
\ket{\nu_{\rm XX};\eta}\bigr> \otimes
\ket{(r,s)_{\rm XX};\eta}\bigr> \otimes
\ket{Gh_{\rm XX};\eta}\bigr>.
\end{align}
The normalization of $\beta\gamma$ ghost Ishibashi/Cardy states 
is summarized in appendix \ref{ICbg}. 
The boundary states of ZZ branes are obtained 
by replacing the Liouville wave functions, for example, 
\begin{align}
S^{(L)}_{\sigma_{\rm NS}}{}^{\nu_{\rm NS}}
=\bigl(2\cosh(2\pi i\nu\sigma) \bigr)
\quad \to \quad
S^{(L)}_{(n,m)_{\rm NS}}{}^{\nu_{\rm NS}}
=\bigl(4\sinh(\pi nb\nu)\sinh(\pi m\nu/b)\bigr).
\end{align}
The normalization of $\eta=+1$ boundary states 
is the same as that of $\eta=-1$ boundary states.%
\footnote{
This should be compared with 
\eq{generalCardy}, 
where the normalization of $\eta=+1$ branes 
is $\sqrt 2$ times bigger than that of $\eta=-1$ branes.}
It is due to the fact that 
open-channel spin fields $\Theta_\pm^{(L+M)}(z)$ 
form a doublet under the combination of 
each Liouville and matter spin-field doublet,
$\Theta_\pm^{(L)}(z)$ and $\Theta_\pm^{(M)}(z)$.%
\footnote{From this fact, the relation 
$n_{ij}{}^k=\mathcal N_{ij}{}^k$ holds in the superstring case. } 
The case of matter Ramond ground states is also the same (the doublet is 
supplied from Liouville spin fields). 

From this expression, we can identify 
the corresponding principal FZZT branes \cite{SeSh}. 
For $\eta=-1$ FZZT branes, 
we can show the relations modulo BRST, 
\begin{align}
\ket{\sigma;(k,l)_{\rm NS_\pm};\eta=-1} = 
\sum_{m,n}\ket{\sigma+\frac{i}{2}(mb-nb^{-1})
       ;{\eta=-1,{\pm(-1)^{\frac{(k-l)-(m-n)}{2}}}}}, \label{m1brst}
\end{align}
with $\ket{\sigma;\eta=-1,\pm}\equiv \ket{\sigma;(1,1)_{\rm NS_\pm};\eta=-1}$ 
following the arguments given in \cite{SeSh,KOPSS}. 
For $\eta =+1$ FZZT branes, on the other hand, there is not 
such a principal brane among the above $\eta=+1$ FZZT branes. Instead, 
we actually show that the following $\eta=+1$ FZZT brane,
\begin{align}
\ket{\sigma;\eta=+1,\pm}\equiv (-1)^{f_R} \ket{\sigma;\eta=-1,\pm}, 
\label{princetap1}
\end{align}
plays a role of a principal $\eta=+1$ FZZT brane:%
\footnote{Note that, in the Liouville wave functions 
of R-R sector, ``$\sinh$'' 
turns to be ``$\cosh$'' 
(``$\cosh$'' turns to be ``$\sinh$'' if one considers $\mu<0$)
in this principal $\eta=+1$ FZZT branes. }
\begin{align}
\ket{\sigma;(k,l)_{\rm R_\pm};\eta=+1} = 
\sum_{m,n}\ket{\sigma+\frac{i}{2}(mb-nb^{-1});\eta=+1,
       {{\pm(-1)^{\frac{(k-l)-(m-n)}{2}}}}}. \label{p1brst}
\end{align}
Because this principal $\eta=+1$ FZZT brane is not a Cardy state 
for the principal $\eta=-1$ FZZT branes, 
these principal branes cannot exist at the same time. 
Even though there is no open string spectrum that propagates 
among these two branes,
they are necessary for the construction of all 
the spectrum of D-branes. 
Since these are simply related 
under the simple transformation $(-1)^{f_R}$, 
it is reminiscent of order/disorder parameters 
in the Kramers-Wanniers duality \cite{KWdual}.%
\footnote{Note that this transformation is not the same 
as that of $\mu\to-\mu$ (or $\eta^{(L)}\to-\eta^{(L)}$).}
Actually we argue in section 3 
that the corresponding annulus amplitudes are not mutually local 
in spacetime. 

The boundary states for type 0A theory are \\
\noindent\underline{\bf Type 0A}
\begin{align}
&\ket{\sigma,(k,l)_{\rm NS};\eta=-1}= \nn\\
&\quad=\int_0^\infty d\nu \, \sum_{(r,s)\in NSNS} 
\bigl(2\cosh(2\pi i\nu\sigma)\bigr)\,
S^{(M)}_{(k,l)}{}^{(r,s)} 
A_{NS}(\nu,(r,s)) \ket{\nu,(r,s);{\rm NSNS};\eta=-1}\bigr>, \nn\\
& \hspace{4cm} (k+l\in 2\mathbb Z), 
\label{aFZZTem1} \\
&\ket{\sigma,(k,l)_{\rm R};\eta=+1}=\nn\\
&\quad=\int_0^\infty d\nu \, \sum_{(r,s)\in NSNS} 
\bigl(2\cosh(2\pi i\nu\sigma)\bigr)\,
S^{(M)}_{(k,l)}{}^{(r,s)} 
A_{NS}(\nu,(r,s)) \ket{\nu,(r,s);{\rm NSNS};\eta=+1}\bigr>, \nn\\
& \hspace{4cm} 
(k+l\in 2\mathbb Z+1), 
\label{aFZZTep1} \\
&\ket{\sigma;(\hat p,\hat q)_{\rm R_\pm};\eta=+1}= \nn\\
&\quad =\frac{1}{2}\int_0^\infty d\nu \, 
\sum_{(r,s)\in NSNS} 
\bigl(2\cosh(2\pi i\nu\sigma)\bigr)S^{(M)}_{(k,l)}{}^{(r,s)} 
A_{NS}(\nu,(r,s))  \ket{\nu,(r,s);{\rm NSNS};\eta=+1}\bigr>
 \,\pm \nn\\
&\qquad\qquad \!\pm \frac{1}{\sqrt 2}\int_0^\infty d\nu \,
\bigl(2i\sinh(2\pi i\nu\sigma+\frac{\epsilon \pi i}{2})\bigr) 
A_{R}(\nu,(\hat p,\hat q)) 
\ket{\nu,(\hat p,\hat q);{\rm RR};\eta=+1}\bigr>. 
\label{aFZZTcgd} 
\end{align}
The first two kinds of boundary states are given as%
\footnote{This normalization 
gives the natural oscillator algebra, 
$[\alpha_n^{[0]},\alpha_m^{[0]}]=n\delta_{n+m}$, 
in the corresponding string field formulation \cite{fi1}. }
\begin{align}
\ket{\sigma,(k,l)_{\rm X}}^{(\rm 0A)}_{\mu>0}=\frac{1}{\sqrt 2}
\Bigl(
\ket{\sigma,(k,l)_{\rm X_+}}^{(\rm 0B)}_{\mu>0}
+
\ket{\sigma,(k,l)_{\rm X_-}}^{(\rm 0B)}_{\mu>0}
\Bigr).
\end{align}
The last boundary state 
of the Ramond ground state is defined by the following 
fusion rule:
\begin{align}
n_{(\frac{p}{2},\frac{q}{2})_+,(\frac{p}{2},\frac{q}{2})_\pm}{}^{(r,s)}
=\frac{1}{2}\, \mathcal 
N_{(\frac{p}{2},\frac{q}{2})_+,(\frac{p}{2},\frac{q}{2})_\pm}{}^{(r,s)}
=\frac{1}{2}\sum_{n=1,2,\cdots,p-1;\ m=1,3,\cdots,q-1}
\delta_{(n,m)_\pm}{}^{(r,s)},
\end{align}
with the identification $(r,s)\sim (p-r,q-s)$.
Under this normalization, 
we actually obtain the following nontrivial 
identification in the pure-supergravity case, $(p,q)=(2,4)$: 
\begin{align}
\ket{\sigma,(1,1)_{\rm NS_\pm};\eta=-1}_{\mu>0}^{\rm (0B)}\ 
&\simeq\  \ket{\sigma,(1,2)_{\rm R_\pm};\eta=+1}_{\mu<0}^{\rm (0A)}, \nn\\
\ket{\sigma,(1,2)_{\rm R}\eta=+1}_{\mu>0}^{\rm (0B)}\ 
&\simeq\  \ket{\sigma,(1,1)_{\rm NS};\eta=-1}_{\mu<0}^{\rm (0A)},
\end{align}
argued in \cite{UniCom,SeSh,SeSh2}. 

Although the following branes are 
the principal FZZT branes of type 0A theory:
\begin{align}
\ket{\sigma;\eta=-1}\equiv \ket{\sigma,(1,1)_{\rm NS};\eta=-1},
\qquad \ket{\sigma;\eta=+1}\equiv (-1)^{f_R} \ket{\sigma;\eta=-1},
\end{align}
the Cardy state of the Ramond ground state, 
$\ket{\sigma;(\hat p,\hat q);\eta=+1}$, cannot be written with 
the above principal FZZT branes. So we must treat these things 
separately. 

\section{Annulus amplitudes of the principal FZZT branes \label{ann}}

In this section, we evaluate the annulus amplitudes of FZZT 
branes. Annulus amplitudes of ZZ branes 
are not considered here, 
since this kind of amplitude is obtained from  
those of the principal FZZT branes 
(with the relation between ZZ and FZZT branes \cite{Mar} 
in this case \cite{SeSh}).

Annulus amplitudes from the CFT approach has been studied in 
\cite{Mar, KOPSS} (bosonic cases) and \cite{SeSh2,Oku} (fermionic case). 
Our technical procedure follows them, and we will 
not repeatedly write such a thing. 
For the later convenience, we denote our amplitudes as 
\begin{align}
Z_{\xi\xi'}^{\eta,\eta'}(&\sigma,(k,l);\sigma',(k',l'))\equiv \nn\\
&\equiv \int_0^\infty dt\,\bra{\sigma,(k,l);\eta=-1,\xi}
\rmq^{\frac{1}{2}(L_0-\frac{c}{24})}
{\tilde\rmq}^{\frac{1}{2}(\bar L_0-\frac{c}{24})}
\ket{\sigma',(k',l');\eta=-1,\xi'} \nn\\
&\equiv Z_{\rm NSNS}^{\eta,\eta'}(\sigma,(k,l);\sigma',(k',l')) 
+\xi\xi' Z_{\rm RR}^{\eta,\eta'}(\sigma,(k,l);\sigma',(k',l')),
\end{align}
with $\rmq=e^{-2\pi t}$.
Throughout this analysis, we use the following relations 
that can be shown with the technique of \cite{cardy2}:
\begin{align}
&\frac{2}{\sqrt{pq}}\sum_{m\in \mathbb Z} \rmq^{\frac{m^2}{2pq}}
\sin(\frac{km}{p}\pi)\sin(\frac{lm}{q}\pi) \nn\\
&\quad =
\left\{
\begin{array}{ll}
\displaystyle
\sum_{(r,s)\in \rm NSNS}\sum_{n\in \mathbb Z}
S^{(M)}_{(k,l)}{}^{(r,s)}
\bigl( \rmq^{\frac{(2npq+qr-ps)^2}{8pq}}
- \rmq^{\frac{(2npq+qr+ps)^2}{8pq}}\bigr), 
& (k-l\in 2\mathbb Z)  \\
\displaystyle
\sum_{(r,s)\in \rm NSNS}\sum_{n\in \mathbb Z}
S^{(M)}_{(k,l)}{}^{(r,s)}
(-1)^{npq}\bigl( \rmq^{\frac{(2npq+qr-ps)^2}{8pq}}
- (-1)^{rs}\rmq^{\frac{(2npq+qr+ps)^2}{8pq}}\bigr), 
& (k-l\in 2\mathbb Z+1), 
\end{array}\right. \nn\\
&\frac{2}{\sqrt{pq}}\sum_{m\in \mathbb Z+pq/2} \rmq^{\frac{m^2}{2pq}}
\sin(\frac{km}{p}\pi+\frac{k}{2}\pi)
\sin(\frac{lm}{q}\pi+\frac{k}{2}\pi) \nn\\
&\quad =\sum_{(r,s)\in \rm RR}\sum_{n\in \mathbb Z}
S^{(M)}_{(k,l)}{}^{(r,s)}
\bigl( \rmq^{\frac{(2npq+qr-ps)^2}{8pq}}
- \rmq^{\frac{(2npq+qr+ps)^2}{8pq}}\bigr), \qquad
(k-l\in 2\mathbb Z) \label{tNSmod}
\end{align}
and the following useful formula that was used in \cite{Mar}:
\begin{align}
\sum_{m\in \mathbb Z}\frac{\cos(Am\pi)}{m^2+\nu^2} = 
\frac{\pi}{\nu}\frac{\cosh\bigl((1-A)\nu\pi\bigr)}{\sinh(\pi \nu)} 
\label{poleeven} \\
\sum_{m\in \mathbb Z+1/2}\frac{\cos(Am\pi)}{m^2+\nu^2} = 
\frac{\pi}{\nu}\frac{\sinh\bigl((1-A)\nu\pi\bigr)}{\cosh(\pi \nu)}. 
\label{poleodd}
\end{align}

\subsection{NSNS exchange amplitudes between FZZT branes}
We now consider the general amplitudes of 
$Z_{\rm NSNS}^{\eta,\eta'}(\sigma,(k,l);\sigma',(k',l'))$, so 
NSNS exchanges. 
This kind of amplitudes does not essentially depend on $\eta$ 
and $\sgn(\mu)$. 
The relation between 0A and 0B is the following:
\begin{align}
&Z_{\rm NSNS}^{(0A)\,\eta,\eta'}(\sigma,(k,l);\sigma',(k',l'))
=2
Z_{\rm NSNS}^{(0B)\,\eta,\eta'}(\sigma,(k,l);\sigma',(k',l')),\quad
((k,l),(r,s)\neq (p/2,q/2))  \nn\\
&Z_{\rm NSNS}^{(0A)\,\eta,+1}(\sigma,(k,l);\sigma',(p/2,q/2))
=
Z_{\rm NSNS}^{(0B)\,\eta,+1}(\sigma,(k,l);\sigma',(p/2,q/2)),\quad
((k,l)\neq (p/2,q/2)) \nn\\
&Z_{\rm NSNS}^{(0A)\,+1,+1}(\sigma,(p/2,q/2);\sigma',(p/2,q/2))
=\frac{1}{2}
Z_{\rm NSNS}^{(0B)\,+1,+1}(\sigma,(p/2,q/2);\sigma',(p/2,q/2)).
\end{align}
So we only consider 0B theory,  
$Z_{\rm NSNS}^{\rm (0B)\,\eta,\eta'}(\sigma,(k,l);\sigma',(k',l'))$. 
Then 0B amplitudes we consider here are 
$(i)$ NSNS amplitudes of $\eta\eta'=+1$ and 
$(ii)$ NSNS amplitudes of $\eta\eta'=-1$. 
They can be expressed as follows: 
\begin{align}
(i)\quad 
&Z_{\rm NSNS}^{\rm (0B)\,\eta,\eta}(\sigma,(k,l);\sigma',(k',l'))\nn\\
&\qquad=\frac{1}{2}\int_0^\infty d\nu 
  \frac{\cosh(2\pi i\nu \sigma)\cosh(2\pi i\nu\sigma')}
       {\sinh(\pi \nu b)\sinh(\pi\nu/b)} 
\sum_{(r,s)\in NSNS}
\dfrac{S^{(M)}_{(k,l)}{}^{(r,s)}S^{(M)}_{(k',l')}{}^{(r,s)}}
      {S^{(M)}_{(1,1)}{}^{(r,s)}}
\times \nn\\
  &\qquad \quad \times
\int_0^\infty dt
        \sum_{n\in \mathbb Z}
\rmq^{\frac{\sigma^2}{2}}(\rmq^{\frac{(2npq+qr-ps)^2}{8pq}}
-\rmq^{\frac{(2npq+qr+ps)^2}{8pq}}) \\
(ii)\quad
&Z_{\rm NSNS}^{\rm (0B)\,\eta,-\eta}(\sigma,(k,l);\sigma',(k',l'))\nn\\
&=-\frac{1}{2}\int_0^\infty d\nu 
  \frac{\cosh(2\pi i\nu \sigma)\cosh(2\pi i\nu\sigma')}
       {\sinh(\pi \nu b)\sinh(\pi\nu/b)} 
\sum_{(r,s)\in NSNS}
\dfrac{S^{(M)}_{(k,l)}{}^{(r,s)}S^{(M)}_{(k',l')}{}^{(r,s)}}
      {S^{(M)}_{(1,1)}{}^{(r,s)}}
\times \nn\\
&\qquad \times \int_0^\infty dt \sum_{n\in \mathbb Z}
(-1)^{npq}\rmq^{\frac{\sigma^2}{2}}(\rmq^{\frac{(2npq+qr-ps)^2}{8pq}}
-(-1)^{rs}\rmq^{\frac{(2npq+qr+ps)^2}{8pq}})
\end{align}
Although they seem to be different forms, 
they come to be a unified form. 
We can actually reexpress the basic amplitudes 
$Z_{\rm NSNS}^{\eta,\eta}(\sigma,(k,l);\sigma',(1,1))$ 
by using \eq{tNSmod} and 
\eq{poleeven} as follows:
\begin{align}
&Z_{\rm NSNS}^{\eta,-1}(\sigma,(k,l);\sigma',(1,1))\nn\\
&=-\eta\int_0^\infty d\nu 
  \frac{\cosh(2\pi i\nu \sigma)\cosh(2\pi i\nu\sigma')}
       {\sinh(\pi \nu b)\sinh(\pi\nu/b)} 
  \times \frac{\sqrt{pq}}{\pi} \times \sum_{m\in \mathbb Z}
        \frac{\sin(\dfrac{km}{p}\pi)\sin(\dfrac{lm}{q}\pi)}
             {pq\,\nu^2+m^2} \nn\\
&=\frac{-\eta}{2}\int_{-\infty}^\infty \frac{d\nu}{\nu}\, 
\frac{\cosh\bigl(2\pi i\dfrac{\nu\sigma}{\sqrt{pq}}\bigr)
      \cosh\bigl(2\pi i\dfrac{\nu\sigma'}{\sqrt{pq}}\bigr)
      }{
      \sinh(\pi\nu/p)\sinh(\pi \nu/q)
      }
\frac{\sinh\bigl((\dfrac{p-k}{p})\pi \nu\bigr)
      \sinh\bigl(\dfrac{l}{q}\pi \nu\bigr)
      }{
      \sinh(\pi\nu)
      }.
\end{align}
By using the fusion relations for $r-s\in 2 \mathbb Z$:
\begin{align}
\frac{S^{(M)}_{(k,l)}{}^{(r,s)}\cdot S^{(M)}_{(k',l')}{}^{(r,s)}}
{S^{(M)}_{(1,1)}{}^{(r,s)}}
= \sum_{m=k+k'-1,k+k'-3,\cdots k-k'+1;
        \atop{n=l+l'-1,l+l'-3,\cdots l-l'+1}}
   S^{(M)}_{(n,m)}{}^{(r,s)},
\end{align}
we obtain the following general formula
\begin{align}
&Z_{\rm NSNS}^{\eta,\eta'}(\sigma,(k,l);\sigma',(k',l'))\nn\\
&=\frac{\eta\eta'}{2}\int_{-\infty}^\infty \frac{d\nu}{\nu}\, 
\frac{\cosh\bigl(2\pi i\dfrac{\nu\sigma}{\sqrt{pq}}\bigr)
      \cosh\bigl(2\pi i\dfrac{\nu\sigma'}{\sqrt{pq}}\bigr)
      }{
      \bigl(\sinh(\pi\nu/p)\sinh(\pi \nu/q)\bigr)^2
      }
\frac{\sinh\bigl((\dfrac{p-k}{p})\pi \nu\bigr)
      \sinh\bigl(\dfrac{l}{q}\pi \nu\bigr)
      \sinh\bigl(\dfrac{k'}{p}\pi \nu\bigr)
      \sinh\bigl(\dfrac{l'}{q}\pi \nu\bigr)
      }{
      \sinh(\pi\nu)
      }.
\end{align}

Following the arguments of \cite{KOPSS}, 
we rewrite this amplitude as
\begin{align}
&=-\frac{\eta\eta'}{2}\int_{-\infty}^\infty \frac{d\nu}{\nu}\, 
\cosh\bigl(2\pi i\dfrac{\nu\sigma}{\sqrt{pq}}\bigr)
      \cosh\bigl(2\pi i\dfrac{\nu\sigma'}{\sqrt{pq}}\bigr)
      \frac{\sinh\bigl(\dfrac{k}{p}\pi \nu\bigr)
      \sinh\bigl(\dfrac{l}{q}\pi \nu\bigr)
      \sinh\bigl(\dfrac{k'}{p}\pi \nu\bigr)
      \sinh\bigl(\dfrac{l'}{q}\pi \nu\bigr)
      }{
      \bigl(\sinh(\pi\nu/p)\sinh(\pi \nu/q)\bigr)^2
      }\times \nn\\
&\quad\qquad\times \biggl\{
\frac{\cosh(\pi\nu)}{\sinh(\pi\nu)}-\frac{\cosh(k\pi\nu/p)}{\sinh(k\pi\nu/p)}
\biggr\}
\end{align}
The first term in the parenthesis 
is the main part of this amplitude, since 
the second term is actually a contribution 
from the unphysical poles of NSNS sector, 
$\nu =inp \, \ (n\in \mathbb Z)$,%
\footnote{It can be easily seen by recalling the correspondence 
with differential operator $\bP^{2n}=(\bsigma \bL^{\hat p})^{2n}$ of
2-component KP hierarchy \cite{fi1}. 
From the viewpoints of string field formulation \cite{fy1,fy2,fy3,fi1}, 
this contribution comes from the normal ordering 
with respect to ${\rm SL}(2,\mathbb C)$ invariant vacuum of $\zeta$ 
\cite{fy1}.}
as is in the bosonic case \cite{KOPSS}. 
The main part can be written with the amplitudes 
of the principal $\eta=\pm 1$ FZZT branes 
following the rule of \eq{m1brst} and \eq{p1brst}:
\begin{align}
&Z_{\rm NSNS,main}^{\eta,\eta'}(\sigma,(k,l);\sigma',(k',l')) =\nn\\
&\qquad =\sum_{m,n;\,m',n'}Z_{\rm NSNS,main}^{\eta,\eta'}
(\sigma+\frac{i}{2}(mb-n/b);\sigma'+\frac{i}{2}(m'b-n'/b)),
\end{align}
where the above basic amplitudes satisfy
\begin{align}
Z_{\rm NSNS,main}^{\eta,\eta}(\sigma;\sigma')
=Z_{\rm NSNS,main}^{-\eta,-\eta}(\sigma;\sigma')
=-Z_{\rm NSNS,main}^{\eta,-\eta}(\sigma;\sigma'). \label{}
\end{align}

\

The amplitudes of $\eta\eta'=+1$ principal FZZT branes
can be evaluated with the procedure of \cite{KOPSS} as 
\begin{align}
Z_{\rm NSNS}^{\eta,\eta}(\sigma;\sigma')
&=\frac{1}{2}\ln\biggl(
         \frac{\sinh(\pi\dfrac{\sigma+\sigma'}{\sqrt{pq}})
               \sinh(\pi\dfrac{\sigma-\sigma'}{\sqrt{pq}})}
              {\sinh(p\pi\dfrac{\sigma+\sigma'}{\sqrt{pq}})
              \sinh(p\pi\dfrac{\sigma-\sigma'}{\sqrt{pq}})}
               \biggr) \nn\\
&=\frac{1}{2}\ln\biggl(
         \frac{\cosh(2\pi\dfrac{\sigma}{\sqrt{pq}})
              -\cosh(2\pi\dfrac{\sigma'}{\sqrt{pq}})}
              {\cosh(2\pi\dfrac{p\sigma}{\sqrt{pq}})
              -\cosh(2\pi\dfrac{p\sigma'}{\sqrt{pq}})}
               \biggr).
\end{align}
At that time, we should be careful to treat 
the boundary cosmological constants 
$\zeta$ \cite{fuku-hoso} and 
the uniformization parameter $z$ \cite{SeSh}. 
We now define these parameters as 
\begin{align}
\tau\equiv \pi \frac{\sigma}{\sqrt{\hat p\hat q}},\qquad 
z = \cosh\tau, \qquad 
\zeta= 
\left\{
\begin{array}{ll}
  \sqrt{|\mu|}\cosh(\hat p \tau) & (\hat \eta=-1) \\
  \sqrt{|\mu|}\sinh(\hat p \tau) & (\hat \eta=+1)
\end{array}\right.,
\end{align}
where we use the label $(\hat p,\hat q)$ 
that is more suitable from the point of view of 
two-matrix models (or 2-component KP hierarchy) \cite{fi1}:
\begin{align}
(\hat p,\hat q)&=(p/2,q/2) \quad (\hbox{even model}), \nn\\
 &=(p,q)\quad (\hbox{odd model}).
\end{align} 
Therefore we obtain the following form: 
\begin{align}
Z_{\rm NSNS}^{\eta,\eta}(\sigma;\sigma')
=
\left\{ 
\begin{array}{ll}
\dfrac{1}{2}\ln\biggl(
         \dfrac{z-z'}
              {\zeta^2-\zeta'^2}
               \biggr) & (\hbox{even model}) \\
\dfrac{1}{2}\ln\biggl(
         \dfrac{z^2-z'^2}
              {\zeta^2-\zeta'^2}
               \biggr) & (\hbox{odd model}).
\end{array} \right.
\end{align}
Note that the dependence of $\zeta $ is changed as 
\begin{align}
z
=\left\{
\begin{array}{ll}
\dfrac{1}{2}\Bigl((\zeta+\sqrt{\zeta^2-\mu})^{1/\hat p} 
+ (\zeta-\sqrt{\zeta^2-\mu})^{1/\hat p} \Bigr) \equiv z(\zeta)
& (\hat \eta=-1), \\
\dfrac{1}{2}\Bigl((\sqrt{\zeta^2+\mu} +\zeta)^{1/\hat p}
+(\sqrt{\zeta^2+\mu} -\zeta)^{1/\hat p}\Bigr) \equiv \tilde z(\zeta) 
& (\hat \eta=+1).
\end{array}\right.
\end{align}
That is, $\zeta$ of $\eta=\pm 1$ FZZT branes are simply related as 
$\zeta_{\eta=-1}^2-\zeta_{\eta=+1}^2=\mu$.

The principal FZZT brane amplitudes of $\eta\eta'=-1$
are somewhat more complicated but we can say that 
the main parts are given as  
\begin{align}
Z_{\rm NSNS,main}^{\eta,-\eta}(\sigma;\sigma')
=
\left\{ 
\begin{array}{ll}
-\dfrac{1}{2}\ln\bigl(
         {z-z'}
               \bigr) & (\hbox{even model}) \\
-\dfrac{1}{2}\ln\bigl(
         {z^2-z'^2}
               \bigr) & (\hbox{odd model}).
\end{array} \right.,
\end{align}
from the modulo-BRST equations \eq{p1brst}. 
This means that 
if we have both brane operators $\psi_\eta(z)$ 
of $\eta=\pm1$ principal branes, 
they should have the following behavior:%
\footnote{Although we do not know the form of unphysical parts, 
these contributions are expected to be nothing but the normal ordering 
with respect to 
${\rm SL}(2,\mathbb C)$ invariant vacuum of $\zeta$ \cite{fy1}. }
\begin{align}
\psi_{\eta=+1}(z')\,\psi_{\eta=-1}(z) \sim (z'-z)^{-1/2},
\end{align} 
since there are no RR exchange between these branes. 
In this sense, the principal $\eta=-1/+1$ branes are not mutually local 
in $z$ (or $\zeta$) spacetime coordinate.%
\footnote{For the relations between a complex coordinate $\zeta$ 
and the 2-dimensional spacetime coordinate $(\phi_L,X_{\rm FF})$
of minimal string theory, see \cite{fy3}
and \cite{MMSS}. } 
That is, their square-root cut cannot dissolve in 
the asymptotic weak coupling region, $z\to \infty$ (or $\zeta\to \infty$). 
Note that this is due to the breaking of the Cardy consistency conditions. 
Actually the Cardy states of $\eta=+1$ FZZT branes 
do not have such a behavior, 
because they are linear combinations 
of the principal $\eta=+1$ FZZT branes, the number of which is even,
and then the square root dissolves in the weak coupling region, 
$z\to \infty$ (or $\zeta\to \infty$). 
So we conclude that 
the principal $\eta=\pm1$ FZZT branes in minimal superstring theory 
can be interpreted as order/disorder parameters in superstring spacetime.

\subsection{RR exchange amplitudes between FZZT branes}
Next we consider RR exchange amplitudes. 
Note that annulus amplitudes between $\eta=-1$ and 
$\eta=+1$ branes must vanish, $Z_{\rm RR}^{\eta,-\eta}=0$. 
It is because the superconformal residual symmetry is 
remained in cylinder and we need to insert a vertex 
operator to obtain non-zero results. 
Thus we now neglect this contribution. 
The amplitudes we consider are 
$(iii)$ RR-ground state exchange amplitudes 
in 0A theory, 
$(iv)$ RR amplitudes of $\eta=\mp1$. 
First the case of $(iii)$ is written as 
\begin{align}
(iii)\quad 
&Z_{\rm RR}^{+1,+1}
(\sigma,(\frac{p}{2},\frac{q}{2});\sigma',(\frac{p}{2},\frac{q}{2}))\nn\\
&\qquad=-\frac{1}{2}\int_0^\infty d\nu 
  \frac{\sinh(2\pi i\nu \sigma+\dfrac{\epsilon}{2}\pi i)
        \sinh(2\pi i\nu\sigma'+\dfrac{\epsilon}{2}\pi i)}
       {\cosh(\pi \nu b)\cosh(\pi\nu/b)} \times \nn\\
&\qquad \qquad \qquad \qquad  \times 
\frac{\sqrt{pq}}{4}\int_0^\infty dt 
       \sum_{n\in \mathbb Z} 
\rmq^{\frac{\nu^2}{2}}(\rmq^{\frac{(2npq)^2}{8pq}}
-\rmq^{\frac{(2npq+pq)^2}{8pq}}) 
\qquad\qquad\qquad\quad\nn\\
&\qquad=-\frac{1}{2}\int_{-\infty}^\infty \frac{d\nu}{\nu} 
  \frac{\sinh(2\pi i \dfrac{\nu \sigma}{\sqrt{pq}}+\dfrac{\epsilon}{2}\pi i)
        \sinh(2\pi i \dfrac{\nu\sigma'}{\sqrt{pq}}+\dfrac{\epsilon}{2}\pi i)}
       {4\cosh(\pi \nu /p)\cosh(\pi\nu/q)}
  \frac{1}{\sinh(2\pi\nu/pq)}, 
\end{align}
and the remaining case of $(iv)$ is 
\begin{align}
&Z_{\rm RR}^{-1,-1}(\sigma,(k,l);\sigma',(k',l'))\nn\\
&\quad =-\frac{1}{2}\int_0^\infty d\nu 
  \frac{\cosh(2\pi i\nu \sigma+\dfrac{\epsilon}{2}\pi i)
        \cosh(2\pi i\nu\sigma'+\dfrac{\epsilon}{2}\pi i)}
       {\cosh(\pi \nu b)\cosh(\pi\nu/b)} 
       \sum_{(r,s)\in RR}
\dfrac{S^{(M)}_{(k,l)}{}^{(-r,-s)}S^{(M)}_{(k',l')}{}^{(r,s)}}
      {S^{(M)}_{(1,1)}{}^{(r,s)}} \times \nn\\
&  \qquad \quad  \times\int_0^\infty dt \sum_{n\in \mathbb Z}
\rmq^{\frac{\nu^2}{2}}(\rmq^{\frac{(2npq+qr-ps)^2}{8pq}}
-\rmq^{\frac{(2npq+qr+ps)^2}{8pq}}),\\
&Z_{\rm RR}^{+1,+1}(\sigma,(k,l);\sigma',(k',l'))\nn\\
&\quad =-\frac{1}{2}\int_0^\infty d\nu 
  \frac{\bigl(i\sinh(2\pi i\nu \sigma+\dfrac{\epsilon}{2}\pi i)\bigr)
        \bigl(i\sinh(2\pi i\nu\sigma'+\dfrac{\epsilon}{2}\pi i)\bigr)}
       {\cosh(\pi \nu b)\cosh(\pi\nu/b)} 
       \sum_{(r,s)\in RR}
\dfrac{\psi^{(M)}_{(k,l)}{}^{(-r,-s)}\psi^{(M)}_{(k',l')}{}^{(r,s)}}
      {S^{(M)}_{(1,1)}{}^{(r,s)}} \times \nn\\
&\qquad\quad \times \int_0^\infty dt  \sum_{n\in \mathbb Z}
\rmq^{\frac{\nu^2}{2}}(\rmq^{\frac{(2npq+qr-ps)^2}{8pq}}
-\rmq^{\frac{(2npq+qr+ps)^2}{8pq}}), 
\end{align}
The basic amplitudes 
$Z_{\rm RR}^{-1,-1}(\sigma,(k,l);\sigma',(1,1))\ \ (k+l\in 2\mathbb Z)$ are 
\begin{align}
&Z_{\rm RR}^{-1,-1}(\sigma,(k,l);\sigma',(1,1))\nn\\
&\quad =-\int_0^\infty d\nu 
  \frac{\cosh(2\pi i\nu \sigma+\dfrac{\epsilon \pi i}{2})
        \cosh(2\pi i\nu\sigma'+\dfrac{\epsilon \pi i}{2})}
       {\cosh(\pi \nu b)\cosh(\pi\nu/b)}  \times \nn\\
&\qquad \qquad \qquad \times \frac{\sqrt{pq}}{\pi} 
    \sum_{m\in \mathbb Z+pq/2}
        \frac{\sin(\dfrac{km}{p}\pi+\dfrac{k\pi}{2})
              \sin(\dfrac{lm}{q}\pi+\dfrac{k\pi}{2})}
             {pq\,\nu^2+m^2}  \nn\\
&\quad =
\left\{
\begin{array}{l}
\displaystyle
-\frac{1}{2}\int_{-\infty}^\infty \frac{d\nu}{\nu}\, 
\frac{\cosh\bigl(2\pi i\dfrac{\nu\sigma}{\sqrt{pq}}+\dfrac{\epsilon \pi i}{2}\bigr)
      \cosh\bigl(2\pi i\dfrac{\nu\sigma'}{\sqrt{pq}}+\dfrac{\epsilon \pi i}{2}\bigr)
      }{
      \cosh(\pi\nu/p)\cosh(\pi \nu/q)
      } \times \\
\displaystyle 
\qquad \qquad \qquad \times \frac{(-1)^{\frac{l-k}{2}}
      \sinh\bigl((\dfrac{p-k}{p})\pi \nu+\dfrac{k}{2}\pi i\bigr)
      \sinh\bigl(\dfrac{l}{q}\pi \nu -\dfrac{l}{2}\pi i\bigr)
      }{
      \sinh(\pi\nu)
      } 
\qquad  \hbox{(even model)}     \\
\displaystyle
-\frac{1}{2}\int_{-\infty}^\infty \frac{d\nu}{\nu}\, 
\frac{\cosh\bigl(2\pi i\dfrac{\nu\sigma}{\sqrt{pq}}+\dfrac{\epsilon \pi i}{2}\bigr)
      \cosh\bigl(2\pi i\dfrac{\nu\sigma'}{\sqrt{pq}}+\dfrac{\epsilon \pi i}{2}\bigr)
      }{
      \cosh(\pi\nu/p)\cosh(\pi \nu/q)
      } \times \\
\displaystyle
\qquad \qquad \qquad \times \frac{(-1)^{\frac{l-k}{2}}
      \cosh\bigl((\dfrac{p-k}{p})\pi \nu+\dfrac{k}{2}\pi i\bigr)
      \sinh\bigl(\dfrac{l}{q}\pi \nu -\dfrac{l}{2}\pi i\bigr)
      }{
      \cosh(\pi\nu)
      } \qquad  \hbox{(odd model)} 
\end{array}\right.
\end{align}
By using the fusion relations for $r-s\in 2\mathbb Z+1$,%
\footnote{Notice that 
$(-1)^{\frac{[(k-l)-(k'-l')](r-s)}{2}+1} 
= (-1)^{\frac{[(k+k')-(l+l')](r-s)}{2}}$.}
\begin{align}
\frac{S^{(M)}_{(k,l)}{}^{(-r,-s)}\cdot S^{(M)}_{(k',l')}{}^{(r,s)}}
{S^{(M)}_{(1,1)}{}^{(r,s)}}
= \sum_{m=k+k'-1,k+k'-3,\cdots k-k'+1;
        \atop{n=l+l'-1,l+l'-3,\cdots l-l'+1}}
   (-1)^{\frac{n-m}{2}}
   S^{(M)}_{(n,m)}{}^{(r,s)}, \nn\\
\frac{\psi^{(M)}_{(k,l)}{}^{(-r,-s)}\cdot \psi^{(M)}_{(k',l')}{}^{(r,s)}}
{S^{(M)}_{(1,1)}{}^{(r,s)}}
= \sum_{m=k+k'-1,k+k'-3,\cdots k-k'+1;
        \atop{n=l+l'-1,l+l'-3,\cdots l-l'+1}}
   (-1)^{\frac{n-m}{2}}
   S^{(M)}_{(n,m)}{}^{(r,s)},
\end{align}
we obtain the formula for the $\eta=-1$ case,
\begin{align}
&Z_{\rm RR}^{-1,-1}(\sigma,(k,l);\sigma',(k',l'))\nn\\
&=
\left\{
\begin{array}{l}
\displaystyle
+\frac{1}{2}\int_{-\infty}^\infty \frac{d\nu}{\nu}\, 
\frac{\cosh\bigl(2\pi i\dfrac{\nu\sigma}{\sqrt{pq}}+\dfrac{\epsilon\pi i}{2}\bigr)
      \cosh\bigl(2\pi i\dfrac{\nu\sigma'}{\sqrt{pq}}+\dfrac{\epsilon\pi i}{2}\bigr)
      }{
      \bigl(\cosh(\pi\nu/p)\cosh(\pi \nu/q)\bigr)^2
      } \times \\
\displaystyle
\qquad \times \frac{
      \sinh\bigl((\dfrac{p-k}{p})\pi \nu+\dfrac{k}{2}\pi i\bigr)
      \sinh\bigl(\dfrac{l}{q}\pi \nu -\dfrac{l}{2}\pi i\bigr)
      \sinh\bigl(\dfrac{k'}{p}\pi \nu-\dfrac{k'}{2}\pi i\bigr)
      \sinh\bigl(\dfrac{l'}{q}\pi \nu -\dfrac{l'}{2}\pi i\bigr)
      }{
      \sinh(\pi\nu)
      } \\
\hspace{13cm} \hbox{(even model)}     \\
\displaystyle
+\frac{1}{2}\int_{-\infty}^\infty \frac{d\nu}{\nu}\, 
\frac{\cosh\bigl(2\pi i\dfrac{\nu\sigma}{\sqrt{pq}}+\dfrac{\epsilon\pi i}{2}\bigr)
      \cosh\bigl(2\pi i\dfrac{\nu\sigma'}{\sqrt{pq}}+\dfrac{\epsilon\pi i}{2}\bigr)
      }{
      \bigl(\cosh(\pi\nu/p)\cosh(\pi \nu/q)\bigr)^2
      } \times \\
\displaystyle
\qquad \times
      \frac{
      \cosh\bigl((\dfrac{p-k}{p})\pi \nu+\dfrac{k}{2}\pi i\bigr)
      \sinh\bigl(\dfrac{l}{q}\pi \nu -\dfrac{l}{2}\pi i\bigr)
      \sinh\bigl(\dfrac{k'}{p}\pi \nu-\dfrac{k'}{2}\pi i\bigr)
      \sinh\bigl(\dfrac{l'}{q}\pi \nu -\dfrac{l'}{2}\pi i\bigr)
      }{
      \cosh(\pi\nu)
      } \\
\hspace{13cm} \hbox{(odd model)} 
\end{array}\right.
\end{align}
for the case of $\eta=+1$, 
\begin{align}
&Z_{\rm RR}^{+1,+1}(\sigma,(k,l);\sigma',(k',l'))\nn\\
&=
\left\{
\begin{array}{l}
\displaystyle
+\frac{1}{2}\int_{-\infty}^\infty \frac{d\nu}{\nu}\, 
\frac{i\sinh\bigl(2\pi i\dfrac{\nu\sigma}{\sqrt{pq}}+\dfrac{\epsilon\pi i}{2}\bigr) \cdot
      i\sinh\bigl(2\pi i\dfrac{\nu\sigma'}{\sqrt{pq}}+\dfrac{\epsilon\pi i}{2}\bigr)
      }{
      \bigl(\cosh(\pi\nu/p)\cosh(\pi \nu/q)\bigr)^2
      } \times \\
\displaystyle
\qquad \times \frac{
      \sinh\bigl((\dfrac{p-k}{p})\pi \nu+\dfrac{k}{2}\pi i\bigr)
      \sinh\bigl(\dfrac{l}{q}\pi \nu -\dfrac{l}{2}\pi i\bigr)
      \sinh\bigl(\dfrac{k'}{p}\pi \nu-\dfrac{k'}{2}\pi i\bigr)
      \sinh\bigl(\dfrac{l'}{q}\pi \nu -\dfrac{l'}{2}\pi i\bigr)
      }{
      \sinh(\pi\nu)
      } \\
\hspace{13cm} \hbox{(even model)}     \\
\displaystyle
+\frac{1}{2}\int_{-\infty}^\infty \frac{d\nu}{\nu}\, 
\frac{i\sinh\bigl(2\pi i\dfrac{\nu\sigma}{\sqrt{pq}}+\dfrac{\epsilon\pi i}{2}\bigr) \cdot
      i\sinh\bigl(2\pi i\dfrac{\nu\sigma'}{\sqrt{pq}}+\dfrac{\epsilon\pi i}{2}\bigr)
      }{
      \bigl(\cosh(\pi\nu/p)\cosh(\pi \nu/q)\bigr)^2
      } \times \\
\displaystyle
\qquad \times
      \frac{
      \cosh\bigl((\dfrac{p-k}{p})\pi \nu+\dfrac{k}{2}\pi i\bigr)
      \sinh\bigl(\dfrac{l}{q}\pi \nu -\dfrac{l}{2}\pi i\bigr)
      \sinh\bigl(\dfrac{k'}{p}\pi \nu-\dfrac{k'}{2}\pi i\bigr)
      \sinh\bigl(\dfrac{l'}{q}\pi \nu -\dfrac{l'}{2}\pi i\bigr)
      }{
      \cosh(\pi\nu)
      } \\
\hspace{13cm} \hbox{(odd model)} 
\end{array}\right.
\end{align}

Also in this case, we can separate the amplitudes into the sum of 
unphysical parts, $\nu=i(2n+1)\hat p$, and main parts. 
The main parts can be written with that of principal FZZT branes as 
\begin{align}
&Z_{\rm RR,main}^{\eta,\eta}(\sigma,(k,l);\sigma',(k',l')) =\nn\\
& =\sum_{n,m;\,n',m'}
(-1)^{\frac{[(k-l)-(n-m)]}{2}+\frac{[(k'-l')-(n'-m')]}{2}}
Z_{\rm RR,main}^{\eta,\eta}
(\sigma+\frac{i}{2}(nb-m/b);\sigma'+\frac{i}{2}(n'b-m'/b)),
\end{align}
and as is noted in section 2.2, \eq{princetap1}, one can find that 
\begin{align}
Z_{\rm RR}^{-1,-1}(\sigma;\sigma')
=Z_{\rm RR}^{+1,+1}(\sigma;\sigma').
\end{align}
The amplitudes of principal FZZT branes are  \\
\noindent\underline{\bf (even model)}
\begin{align}
&Z_{\rm RR}^{-1,-1}(\sigma;\sigma')_{\mu>0} \nn\\
&\quad=\dfrac{1}{2}\ln\bigl(\cosh(\tau)-\cosh(\tau')\bigr)
+\dfrac{1}{2}\ln\biggl(
\dfrac{\cosh(\hat p\tau)+\cosh(\hat p \tau')}
      {\cosh(\hat p\tau)-\cosh(\hat p \tau')}
\biggr) \nn\\
&\quad=\dfrac{1}{2}\ln\bigl(z-z'\bigr) 
+\dfrac{1}{2}\ln\biggl(\dfrac{\zeta+\zeta'}{\zeta-\zeta'}\biggr) \\
&Z_{\rm RR}^{-1,-1}(\sigma;\sigma')_{\mu<0} \nn\\
&\quad=\dfrac{1}{2}\ln\bigl(\cosh(\tau)-\cosh(\tau')\bigr)
-\dfrac{1}{2}\ln\biggl(
\dfrac{\sinh(\hat p\tau)+\sinh(\hat p \tau')}
      {\sinh(\hat p\tau)-\sinh(\hat p \tau')}
\biggr) 
- \ln\bigl(\sinh\bigl(\dfrac{\tau+\tau'}{2}\bigr)\bigr)
\nn\\
&\quad =\dfrac{1}{2}\ln\bigl(z-z'\bigr) 
-\dfrac{1}{2}\ln\biggl(\dfrac{\zeta+\zeta'}{\zeta-\zeta'}\biggr) 
-\dfrac{1}{2}
\ln\Bigl(zz'+
\sqrt{(z^2-1)(z'^2-1)}
-1
\Bigr) 
\end{align}
\noindent\underline{\bf (odd model)}
\begin{align}
&Z_{\rm RR}^{-1,-1}(\sigma;\sigma')_{\mu>0} \nn\\
&\quad =\dfrac{1}{2}\ln\biggl(\dfrac{\cosh(\tau)-\cosh(\tau')}
                           {\cosh(\tau)+\cosh(\tau')}\biggr)
+\dfrac{1}{2}\ln\biggl(
\dfrac{\cosh(\hat p\tau)+\cosh(\hat p \tau')}
      {\cosh(\hat p\tau)-\cosh(\hat p \tau')}
\biggr) \hspace{1cm} \qquad \quad \quad\ \ \ \nn\\
&\quad=\dfrac{1}{2}\ln\biggl(\dfrac{z-z'}{z+z'}\biggr) 
+\dfrac{1}{2}\ln\biggl(\dfrac{\zeta+\zeta'}{\zeta-\zeta'}\biggr) \\
&Z_{\rm RR}^{-1,-1}(\sigma;\sigma')_{\mu<0} \nn\\
&\quad=\dfrac{1}{2}\ln\biggl(\dfrac{\sinh(\tau)-\sinh(\tau')}
                              {\sinh(\tau)+\sinh(\tau')}\biggr)
+\dfrac{1}{2}\ln\biggl(
\dfrac{\sinh(\hat p\tau)+\sinh(\hat p \tau')}
      {\sinh(\hat p\tau)-\sinh(\hat p \tau')}
\biggr) \nn\\
&\quad=\dfrac{1}{2}\ln\biggl(
          \dfrac{\sqrt{z^2-1}-\sqrt{z'^2-1}}
                {\sqrt{z^2-1}+\sqrt{z'^2-1}}\biggr) 
+\dfrac{1}{2}\ln\biggl(\dfrac{\zeta+\zeta'}{\zeta-\zeta'}\biggr) 
\end{align}

\

We then summarize the full amplitudes (${\rm NSNS}+{\rm RR}$) 
of the principal $\eta=-1$ FZZT branes. 
For 0B theory, \\
\noindent\underline{\bf (even model)}
\begin{align}
Z^{-1,-1}_{\xi_1,\xi_2}(\zeta_1;\zeta_2)_{\mu>0} 
&=\dfrac{1+\xi_1\xi_2}{2}
\ln\biggl(\dfrac{z_1-z_2}{\zeta_1-\zeta_2}\biggr)
-\dfrac{1-\xi_1\xi_2}{2}\ln(\zeta_1+\zeta_2)   \\
Z^{-1,-1}_{\xi_1,\xi_2}(\zeta_1;\zeta_2)_{\mu<0}
&=\dfrac{1+\xi_1\xi_2}{2}
\ln\biggl(\dfrac{z_1-z_2}{\zeta_1-\zeta_2}\biggr)
-\dfrac{1-\xi_1\xi_2}{2}\ln(\zeta_1+\zeta_2) - \nn\\
&\quad-\dfrac{\xi_1\xi_2}{2}
\ln\Bigl(
z_1z_2-
\sqrt{(z_1^2-1)(z_2^2-1)}
-1
\Bigr) 
\end{align}
\noindent\underline{\bf (odd model)}
\begin{align}
&Z^{-1,-1}_{\xi_1,\xi_2}(\zeta_1;\zeta_2)_{\mu>0}
=\dfrac{1+\xi_1\xi_2}{2}
\ln\biggl(\dfrac{z_1-z_2}{\zeta_1-\zeta_2}\biggr)
+\dfrac{1-\xi_1\xi_2}{2} 
\ln\biggl(\dfrac{z_1+z_2}{\zeta_1+\zeta_2}\biggr) \\
&Z^{-1,-1}_{\xi_1,\xi_2}(\zeta_1;\zeta_2)_{\mu<0} = \nn\\
&\qquad =\dfrac{1+\xi_1\xi_2}{2}
\ln\biggl(\dfrac{\sqrt{z_1^2-1}-\sqrt{z_2^2-1}}{\zeta_1-\zeta_2}\biggr)
+\dfrac{1-\xi_1\xi_2}{2}
\ln\biggl(\dfrac{\sqrt{z_1^2-1}+\sqrt{z_2^2-1}}{\zeta_1+\zeta_2}\biggr), 
\end{align}
and for 0A theory, 
\begin{align}
Z^{-1,-1}_{\xi_1,\xi_2}(\zeta_1;\zeta_2)
=\left\{
\begin{array}{ll}
\ln\biggl(
         \dfrac{z-z'}
              {\zeta^2-\zeta'^2}
               \biggr) & (\hbox{even model}) \nn\\
\ln\biggl(
         \dfrac{z^2-z'^2}
              {\zeta^2-\zeta'^2}
               \biggr) & (\hbox{odd model}),
\end{array} \right.
\end{align}
are obtained. 

Also for the pure-supergravity case of $(p,q)=(2,4)$, 
(note that $z$ and $\zeta$ accidentally coincide, $z=\zeta=\cosh\tau$),
we can actually show that 
\begin{align}
Z_{\rm NSNS}^{-1,-1}(\sigma_1,(1,1);\sigma_2,(1,1))_{\mu>0}
&=-\frac{1-\xi_1\xi_2}{2}\ln(\zeta_1+\zeta_2), \nn\\
Z_{\rm NSNS}^{-1,-1}(\sigma_1,(1,1);\sigma_2,(1,1))_{\mu<0}
&=-\ln\bigl(\cosh(\frac{\tau_1+\xi_1\xi_2\tau_2}{2})\bigr) \nn\\
&=-\frac{1}{2}\ln\Bigl(\sqrt{(\zeta_1^2-|\mu|)(\zeta_2^2-|\mu|)}
                             +\xi_1\xi_2\,\zeta_1\zeta_2-|\mu|\Bigr).
\end{align}
Of course, 
this is the previous results of this case \cite{Oku}.

\section{Conclusion and discussion}

In this paper, we investigate the explicit form of boundary states in 
$(p,q)$ minimal superstring theory. For this purpose, we 
actually show the way to obtain 
all the wave functions of $\eta=\pm$ Cardy states within the 
modular bootstrap methods in $\mathcal N=1$ superconformal field theory. 
We then identify the corresponding principal 
$\eta=\pm1$ FZZT branes following the arguments given in \cite{SeSh} 
and explicitly evaluate these annulus amplitudes. 
The principal $\eta=-1/+1$ FZZT branes of 0B theory 
are interpreted as order/disorder parameters which causes
the Kramers-Wannier duality in the spacetime sense of this superstring theory. 

From the analysis of \cite{SeSh}, it was realized that 
among many different FZZT branes 
only a few numbers of principal FZZT branes are 
important and they correspond to the fundamental 
degrees of freedom of the theory. 
Since we can extract all the closed-string degrees of freedom 
from the principal $\eta=-1$ FZZT and its anti-FZZT branes \cite{fi1}, 
it is natural to think of the principal $\eta=-1$ 
FZZT brane as independent degrees of freedom. 

In the case of Ising model, however, 
we can clearly see the relation with the fermion system 
by introducing the disorder parameter. 
In this sense, it is interesting if we could find 
some more general structures of minimal superstring theory, 
by considering how to describe the principal $\eta=+1$ FZZT branes 
in the exact nonperturbative formulations. 

Of course this kind of duality 
is very familiar in conformal field theory, as 
the T-duality of worldsheet descriptions.%
\footnote{See \cite{TdualNonCritical} for the investigations 
of the worldsheet Kramers-Wannier duality (T-duality) 
in non-critical string theory .}
A new feature of our Kramers-Wannier like duality 
is that order/disorder parameters in minimal superstring theory 
correspond to D-branes in spacetime (not worldsheet observables). 
Since this structure is originated from the basic nature of 
the NSR formalism, it is interesting to investigate 
what is the spacetime properties of NSR superstring theory.

\section*{Acknowledgments}
The author thanks M.~Fukuma for useful discussions 
and encouragements, and K.~Okuyama, K.~Murakami, T.~Takayanagi 
and M.~Sato for useful conversations. The author also thanks H.~Hata for 
various supports in preparing this paper. 
This work was supported in part by the Grant-in-Aid for 
the 21st Century COE
``Center for Diversity and Universality in Physics'' 
from the Ministry of Education, Culture, Sports, Science 
and Technology (MEXT) of Japan. 
The author is also supported by 
Research Fellowships of the Japan Society for the Promotion 
of Science (JSPS) for Young Scientists (No.\ 18\textperiodcentered 2672). 

\appendix

\section{Summary of the basic modular functions} \label{modfunc}
Here we summarize our convention of the basic characters:
\begin{align}
\chi_0^{(NS)}(\tau)&\equiv \frac{\rmq^{-1/48}}{\eta (\tau)} 
\prod_{n=1}^\infty (1+\rmq^{n-1/2}) 
=\frac{1}{\eta (\tau)} \sqrt{\frac{\theta_3(\tau)}{\eta(\tau)}},  \nn\\
\chi_0^{(\widetilde{NS})}(\tau)&\equiv \frac{\rmq^{-1/48}}{\eta (\tau)} 
\prod_{n=1}^\infty (1-\rmq^{n-1/2}) 
=\frac{1}{\eta (\tau)} \sqrt{\frac{\theta_4(\tau)}{\eta(\tau)}}, \nn\\
\chi_0^{(R)}(\tau)&\equiv \frac{\rmq^{1/24}}{\eta (\tau)} 
\prod_{n=0}^\infty (1+\rmq^{n}) 
=\frac{\sqrt 2}{\eta (\tau)} \sqrt{\frac{\theta_2(\tau)}{\eta(\tau)}}.
\end{align}
with Dedekind $\eta$-function 
$\eta(\tau)\equiv \rmq^{1/24}\prod_{n=1}^\infty (1-\rmq^n)$ 
and $\rmq=e^{2\pi i \tau}$. 
$\theta_a(\tau)$ is the corresponding theta function. 
The modular transformations are 
\begin{align}
\eta(-1/\tau)
&=\sqrt{\dfrac{\tau}{i}}\eta(\tau),\qquad
\theta_3(-1/\tau)
=\sqrt{\dfrac{\tau}{i}}\theta_3(\tau), \nn\\
\theta_4(-1/\tau)
&=\sqrt{\dfrac{\tau}{i}}\theta_2(\tau), \quad
\theta_2(-1/\tau)
=\sqrt{\dfrac{\tau}{i}}\theta_4(\tau).
\end{align}

\section{The Ishibashi/Cardy states of superconformal ghost} \label{ICbg}
It is useful to note about our convention and notation of 
the Ishibashi/Cardy states of 
superconformal ghost \cite{bdystates}. 
Here we especially consider the normalization of $\beta \gamma$ Cardy states. 
It is convenient for $\beta \gamma $ ghost 
to be written in the form of \cite{Ishibashi}. 
We construct them with Watt's technique denoted in \cite{bppz} as
\begin{align}
\ket{\beta\gamma_q;\eta} \bigr>
&= \sum_N \ket{q;N}\otimes U_\eta A\overline{\ket{-2-q;N^*}}
=U_\eta \,
e^{-\sum_{r<0}(\gamma_r\tilde\beta_r+\tilde\gamma_r\beta_r)}
\ket{q}\otimes\overline{\ket{-2-q}} 
\end{align}
where $q$ is a corresponding picture and $\ket{Gh_q;\eta}\bigr>$ 
is defined with the proper Ishibashi state of $bc$ ghost, $\ket{bc}\bigr>$, 
as $\ket{Gh_q;\eta}\bigr>
\equiv \ket{\beta\gamma_q;\eta}\bigr>\ket{bc}\bigr>$.
We use the following hermitian conjugation, 
\begin{align}
\gamma_r^\dagger=-\gamma_{-r}, \qquad 
\beta_r^\dagger=\beta_{-r}, 
\end{align}
and we define the automorphism $U_\eta$ to satisfy
\begin{gather}
U_\eta \tilde \gamma_r U_\eta^{-1} = -i \eta \tilde \gamma_r, \quad 
U_\eta \tilde \beta_r U_\eta^{-1} = i \eta \tilde \beta_r, \quad 
U_\eta^\dagger=-(-1)^{f_R}\cdot U_\eta^{-1}, \quad 
U_{-\eta}=U_\eta(-1)^{f_R}.
\end{gather}
The normalization of states is 
defined from the open string character summed 
over the picture $q$ Hilbert space $\mathcal H^{(q)}$ 
as follows:
\begin{align}
\bigl<\bra{\beta\gamma_{-1};\eta}T(\rmq)\ket{\beta\gamma_{-1};\eta} \bigr>
&=+\frac{\rmq^{1/24}}{\prod_{n\geq 1}(1+\rmq^{n-1/2})^2} \nn\\
&=+\frac{\tilde \rmq^{1/24}}{\prod_{n\geq 1}(1+\tilde \rmq^{n-1/2})^2}
=-\tr_{\mathcal H^{(-1)}_{\beta\gamma}}\bigl[(-1)^{f_R}
\tilde \rmq^{L_0-c/24}\bigr]
\nn\\
\bigl<\bra{\beta\gamma_{-3/2};\eta}
T(\rmq)\ket{\beta\gamma_{-1/2};\eta} \bigr>
&=-\frac{\rmq^{-1/12}}{2\prod_{n\geq 1}(1+\rmq^{n})^2} \nn\\
&=-\frac{\tilde \rmq^{1/24}}{\prod_{n\geq 1}(1-\tilde \rmq^{n-1/2})^2}
=-\tr_{\mathcal H^{(-1)}_{\beta\gamma}}\bigl[
\tilde \rmq^{L_0-c/24}\bigr]
\nn\\
\bigl<\bra{\beta\gamma_{-1};\eta}T(\rmq)(-1)^{f_R}
\ket{\beta\gamma_{-1};\eta} \bigr>
&=-\frac{\rmq^{1/24}}{\prod_{n\geq 1}(1-\rmq^{n-1/2})^2} \nn\\
&=-\frac{\tilde \rmq^{-1/12}}{2\prod_{n\geq 1}(1+\tilde \rmq^{n})^2} 
=-i\tr_{\mathcal H^{(-1/2)}_{\beta\gamma}}\bigl[(-1)^{f_R}
\tilde \rmq^{L_0-c/24}\bigr],
\end{align}
where 
$T(\rmq)=\rmq^{\frac{1}{2}(L_0-c/24)}\bar \rmq^{\frac{1}{2}(\bar L_0-c/24)}$
The normalization of the second equation 
follows that of the first equation.%
\footnote{
If we assume that 
the character in NS sector is expanded as $\tilde \rmq^{1/24}+\cdots$, 
then the character in $\widetilde{\rm NS}$ sector must be expanded 
as $-\tilde \rmq^{1/24}+\cdots$.}
Note that the normalization (or the sign) 
of the third equation is required from 
the spacetime statistics in superstring theory 
(open strings in R sector are fermions) and that 
this negative sign of the character 
can be consistently obtained from the definition 
$\ket{Gh_q;-\eta}\bigr>=(-1)^{f_R}\ket{Gh_q;\eta}\bigr>$ 
of the closed-channel Ishibashi states.

\setlength{\itemsep}{5.\baselineskip}

\end{document}